# DESIGN AND OPTIMIZATION OF BRAKE DISC USING MULTI-OBJECTIVE GENETIC ALGORITHM

**Swapnil Kumar [1], Sundar V Atre [2]**

[1]*Project Engineer, University of Louisville*
*Additive Manufacturing Institute of Science and Technology, United States (E-mail: swapnilsrivastava1997@gmail.com)*
[2]*Professor, University of Louisville*
*Additive Manufacturing Institute of Science and Technology, United States*

**ABSTRACT**: Analysis of brake disc is carried out to depict the behavior of the braking system with respect to bump, droop in endurance, maneuverability, hill Climb & acceleration. Design calculation and analysis have been carried out for the brake disc and subsequently, design calculations have also been carried out for the brake caliper. Structural, thermal, vibrational, computational fluid dynamics and fatigue analysis has been carried out to optimize and validate the performance of disc brakes. The design of experiments has been carried out for the brake disc in order to optimize the performance of the braking system. Ventilated disc brake with an outer diameter of 175 mm has been used and 83 % performance efficiency is achieved after all the proper validations and analysis. Very fine Meshing has been considered for analyzing the disc brake to obtain maximum efficient results. Stainless Steel (SS-410) Material configuration has been considered for disc brake and performance enhancement of ventilated disc brake is carried out using Matlab, Ansys, and Solidworks. The brake disc is going to be deployed as a common brake disc in the rear part of the ATV responsible for providing effective rear wheel locking. Piaggio double piston fixed calipers have satisfied the piston diameter for wheel locking conditions at rear wheels with DOT-4 Brake fluid in the master cylinder to provide effective braking. The rear disc brake was fixed on the gearbox output shaft and a caliper mount is welded on a rear member of the roll cage. A mathematical model has been generated for carrying out Multi-objective genetic algorithm optimization. The newly designed brake disc is optimum in terms of weight, a factor of safety, thermal dissipation, equivalent stress, vibration with enhanced airflow behavior. Converged residual plots have been obtained in computational fluid dynamics simulation by using 2nd-degree order. In order to meet the frequency of rear disc brake to firing frequency of engine, brake disc has been optimized in terms of vibration considering all the parameters.

**KEY WORDS**: Braking System, Computational Fluid Dynamics, Stainless Steel (SS- 410), Disc brake, Braking efficiency.

## 1. Literature Review

Swapnil et al [1] has presented a paper on the braking system for All -Terrian vehicle. Thermal and computational fluid dynamics work has been carried out for the laminar, forced, and normal convection flow. It has been conclude that from the research paper Improvised designed brake disc is better than the initially designed brake disc after analyzing the perspective of light weight and better thermal dissipation. Yang et al. [2] has presented a paper in which they considered a disc brake system under a constant torque and analyzed it by using FEM. The stress and strain have been analyzed under various amplitudes of loading and a linear relationship between the maximum stress or strain and amplitude of loading has been established. The stress at the critical locations in the rotating disc brake obtained was plane, multiaxial and non-proportional under constant mechanical load. It has been concluded from the research paper that stress and strain in the disc are the cyclic functions of the rotation angle in the case when the brake rotor rotates under the constant braking torque. Swapnil et al. [3] has presented a paper on the effective heat transfer dissipation of braking system under the conditions of the laminar and forced flow. It has been concluded from the paper that heat flux obtained in the laminar convection is less as compared to that obtained in forced convection. Swapnil et al. [4] has presented a paper on the design and development of longitudinal vehicle dynamics for All-terrain vehicle based on the development of longitudinal vehicle dynamics to achieve maximum acceleration and deceleration from the transmission and braking system. Belhocine et al. [5] have carried out their research to develop a finite element model of the disc brake assembly to understand the impact of Young's modulus. They have validated their results by using experimented modal analysis and the effect of Young’s modulus variations for the disc brake components on the squeal propensity using a three-dimensional finite element model has been understood. Simulation results demonstrate the fact that instability of the disc brake is sensitive to Young’s modulus variations of the disc brake components and due to the large variations and reduction in the unstable frequencies better squeal performance has been obtained. Wahlström et al. [6] has carried out the experimentation on the testing of disc brake material in consideration to the generation of airborne wear particles and pressure wearing was observed at low brake pressure. The thesis research content showed computational methods which numerically simulate the wear. It has been concluded that Ultrafine (nanosized), fine and coarse airborne wear particles generated from disc brake materials contains metals such as iron, copper, and tin. Belhocine et al. [7] has investigated the thermo-mechanical behavior between pad and rotor disc. It was concluded from the paper that the ventilation system plays great importance in cooling the disc and research work was carried out for von-misses stress and total deformation of the disc. Structural and thermal simulation and modeling of the temperature in the disc brake is carried out by the variation of parameters. These parameters include the cooling mode of the disc and the material of the disc. The brake disc rotor is made of cast iron with high carbon content and the contact surface of the disc receives an entering heat flux. Thermal and mechanical stresses cause crack propagation and fracture in the bowl and also results in the wear of the disc and pads and it results in the increase of temperature, von-mises stress, total deformations of the disc, and contact pressures of the pads. Cao et al. [8] have presented a paper that displayed a technique for modeling automotive disc brake rotors and the prediction of squeal frequencies. The unstable frequencies of the brakes were obtained from a complex-valued, asymmetric eigenvalue formulation. Tehrani et al. [9] has performed thermal analysis on ventilated disc brake made of graded composite material and steel alloy on ABAQUS software. It was seen that the temperature variation and vertical displacement in the FGM disk are much lower than the steel disk. Simulation results show that the material properties of the disk brake exert an essential influence on the surface temperature, Von-Mises stress distribution, and vertical displacement of the brake disc rotor. It has been concluded that temperature variation and vertical displacement in FGM disk is much lower than steel disk, von-mises stress distribution in radial direction grows gradually and FGM brake disk eliminates thermal cracking and wears in localized contact point or hot spots. Mule et al. [10] have focused on rotor strength and lateral deformation and its connection with actual component temperatures. Validation of FEA assumptions and results were discussed considering both the heating and cooling profiles of the rotor. The static structural and thermal analysis has been done to validate the result and was no significant improvement in temperature profile using inboard connection type rotor and coning deformations. Wang et al. [11] published a paper that aims at the modal solution of disc brake rotor concerning rotation. They have calculated the modal parameters of the stationary disc from the FE model using complex eigenvalue analysis. A finite element model of stationary disc brake has been constructed, modal parameters were calculated, and the equivalent modal parameter is extracted and expressed as a function of rotation speed and original stationary status modal parameter. Óskar et al. [12] have presented a thesis work on Brake rotor, hub, and upright design. The thesis work has been published by from School of Science and Engineering Tækni- og verkfræðideild. Calculation of a brake caliper has been carried out and a new Wilwood GP200 brake caliper was selected for all wheels. The designed brake caliper offer enough braking force to support the proper efficient braking system of Formula car and it is lightweight and has low cost along with the objectives like high reliability, proper adjustability lower weight, and lower center of gravity was achieved in the research content for efficient braking. Ali et al. [13] have presented a paper on predictive tools to evaluate braking system performance using a fully coupled thermomechanical finite element model. It has provided us the methodology for thermomechanical analysis of brake rotor and Thermomechanical analysis of disc brake rotor with grey cast iron material properties has been analyzed.

## 2. Design optimization of Rear disc brake

### 2.1 Mathematical Equation

Mathematical Equations has been used to analyse the effective braking system for an Off-road vehicle.

2.1.1 Design optimization

$$Minimize:\ S = \sum_{n=1}^{n=3} \beta(n)x(n) + \sum_{n=4} \beta(n)x^2(n) + \beta(5);$$
$$Minimize:\ T = \sum_{n=1}^{n=3} \gamma(n)x(n) + \sum_{n=4} \gamma(n)x^2(n) + \gamma(5);$$
$$Minimize:\ F = \sum_{n=1}^{n=3} \delta(n)x(n) + \sum_{n=4} \delta(n)x^2(n) + \delta(5);$$

(1)

2.1.2 Energy Transport Equation

$$\frac{\partial(\rho E)}{\partial t} + \nabla \cdot [\mathbf{V}(\rho E + p)] = \nabla \cdot [K_{eff}\nabla T - \sum h_j \boldsymbol{J_J} + \tau_{eff} \cdot \mathbf{V}] + S_h$$

(2)

2.2.3 Convection heat transfer coefficient for laminar flow

If Re > 2.4 x $10^5$ then

$$h = 0.04 \text{ x} \left(\frac{Ka}{Dd}\right) \text{x } Re^{0.8}$$

(3)

2.2.4 Forced Convection Heat Transfer Coefficient

$$Nu = \frac{h \text{ x } D}{k} \qquad Re = \frac{(D \text{ x } v \text{ x } \rho)}{\mu}$$

(4)

### 2.2 Intial design of the brake disc

Disc brake has been modeled in Solidworks. The initial values for Outer radius (Q1) and thickness (Q2) are considered to be 175mm and 6mm respectively. The CAD geometry is meshed using 1 mm sized tetrahedral quadratic elements. SS-410 material has been considered for disc brake.

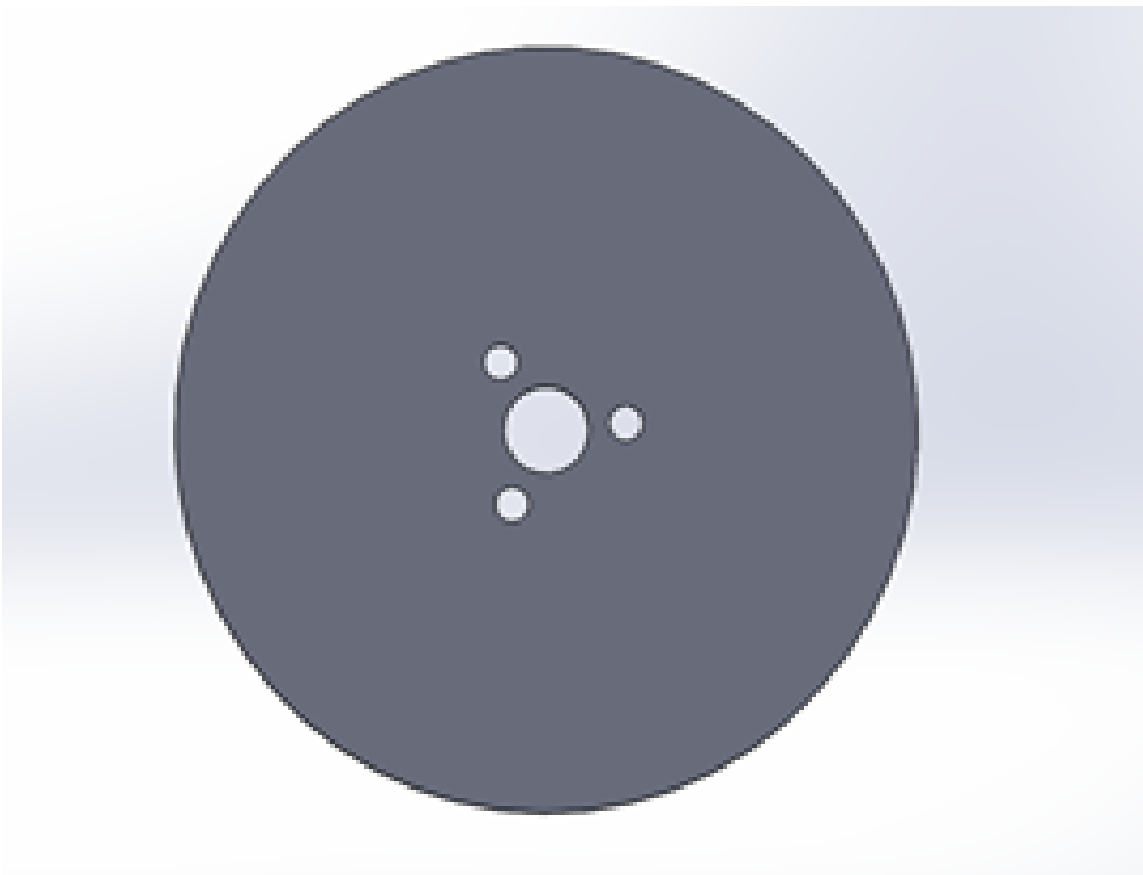

Figure 2. Design

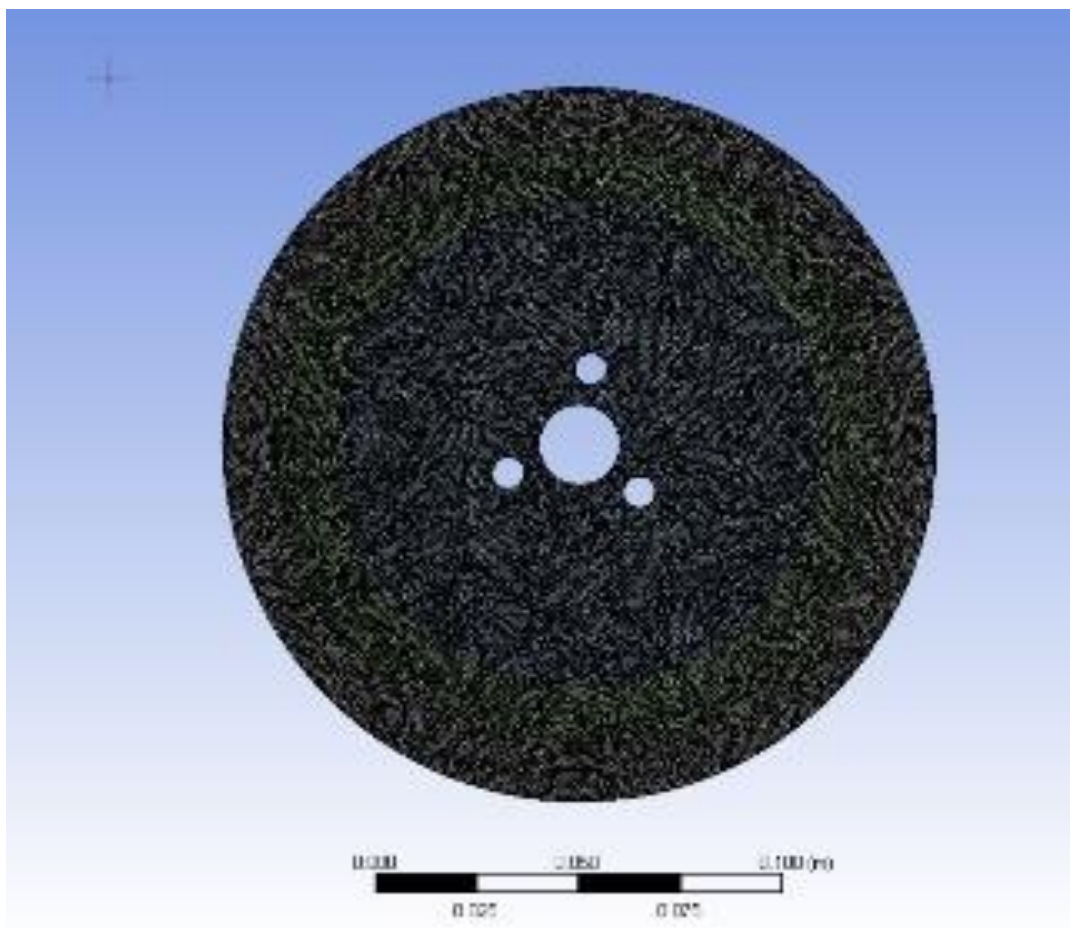

Figure 3. Meshing

Table 1. Meshing

| Nodes | Elements |
|---|---|
| 1784211 | 1238202 |

Table 2. Material properties of SS-410

| Property | Value | Unit |
|---|---|---|
| Density | 7750 | Kg/m^3 |
| Young modulus | 200 | GPa |
| Poisson ratio | 0.280 | - |
| Thermal conductivity | 25 | W/m ◦K |
| Specific heat | 460 | J/kg ◦K |

## 2.3 Design constraints for rear disc brake

Geometrical design constraints for subsystems of rear disc brake. Q1 and Q2 are outer radius and thickness of the disc brake respectively.

| | |
|---|---|
| O1: | $Q1 \geq 159.5$; |
| O2: | $Q1 \leq 190.5$; |
| O3: | $Q2 \geq 5.4$; |
| O4: | $Q2 \leq 6.6$; |

## 2.4 Structural Analysis

For bolting the disc brake 8mm full threaded bolt has been used. Matlab calculation has been done to calculate wheel torque on rear part of vehicle.

Table 3. Vehicle Specifications

| S.No | Parameter | Value |
|---|---|---|
| 1 | Rolling Friction Coefficient | 0.06 |
| 2 | Weight of Vehicle (kg) including Passenger | 310 |
| 3 | Wheelbase (mm) | 1321 |
| 4 | Longitudinal Distance of COG from front axle (mm) | 758 |
| 5 | Longitudinal Distance of COG from rear axle (mm) | 563 |
| 6 | Height of Centre of Gravity (mm) | 547 |
| 7 | Coefficient of friction between road and tyre | 0.67 |
| 8 | Radius of rear wheel (mm) | 152.4 |

Table 4: Structural Brake Parameters

| Denotations | Parameter |
|---|---|
| Kf | Front Brake Force Coefficient |
| Kr | Rear Brake Force Coefficient |
| x1 | distance of CG from front wheels longitudinally |
| x2 | distance of CG from rear wheels longitudinally |
| u | Coefficient of Friction between Road and tire |
| fr | rolling friction coefficient |
| h1 | height of CG from ground |

x1= 75.8; in cm

x2= 56.3; in cm

u = 0.67;

fr = 0.06;

h1 = 54.7; in cm

T1 = Kf;

T2 = Kr;

T1 = x2 + h1*(u+fr);

T2 = x1 – h1*(u+fr);

T3 = T1. /T2;

T3 = 2.68

Kf/Kr = 2.68

Taking Kf/Kr equal to 2.68

Kf + Kr = 1

Kf = 0.73

Kr = 0.27

BF = Total Brake Force

m = mass of Vehicle (Including Passenger)

V = velocity of vehicle

S = Braking Distance

m = 310; in kg

v = 16; in m/sec

S = 3; in m

BF = (1/2 * m * v^ (2)). /S;

BF = 1.3227e+04

Total Braking Force = 1.3227e+04 N

Total Braking force (Rear) = Kr × Braking force (Total)

C1 = Total Braking force (Rear)

C1 = 3571.3 N

Frictional Force on vehicle = u * N;

N = m * g;

g = 9.81;

Frictional Force on vehicle = 2037.54 N

Inertial Force Due to deceleration = m * d;

Since, Frictional Force on vehicle = Inertial Force Due to deceleration

2037.54 = m * d;

d = 6.573;

d/g = 0.67;

Rear axle dynamic load = W - (W*(x2 +[(d/g) *h)])/L;

Rear axle dynamic load = 901.3 N

Frictional torque at rear = Rear axle dynamic load * u * Rr;

Frictional torque at rear = 184 N-m

Frictional Torque for one of rear braking system = 92 N-m

Velocity = pie * Dr * N;

16 = pie *0.175* N

N = 16/(pie*0.175)

Angular velocity of brake disc = 2*pie*N

Angular velocity of brake disc  = 2*pie*(16/(pie*0.175))

Angular velocity of brake disc = 32/0.175

Angular velocity of disc = 183 rad/sec

Table 5. Vehicle obtained parameters

| Parameter | Value |
|---|---|
| Brake Force to lock rear wheel | 3571.3 N |
| Frictional Torque (Rear) | 92 N-m |
| Rotational Velocity | 183 rad/sec |

Table 6. Brake caliper

| S.No | Parameter | Value |
|---|---|---|
| 1 | Master cylinder bore | 18 mm |
| 2 | Area of master cylinder, A(mc) | 2.54 x 10^ (-4) m2 |
| 3 | Area of rear caliper | 698.35 mm2 |
| 4 | Pedal force (PF) | 225 N |
| 5 | Pedal ratio (PR) | 5:1 |
| 6 | Leverage efficiency | 0.8 |

Rear Braking system:

F(mc) = PR * PF * leverage efficiency

P(mc) = F(mc)/A(mc)

F(caliper) = 2* P(mc) * A(caliper) * ηwc

F(caliper) = 2837.44 N

Pad coeffecient = 0.35

Force on rear brake disc = 2*0.35*2837.44

Force on rear brake disc = 1986.2  N

Braking Torque = Braking force * Effective radius of brake disc

Braking Torque = 144 N-m

Table 7. Structural simulation parameters

| Parameter | Value |
|---|---|
| Braking Force (Rear) | 1986.2 N |
| Brake Torque (Rear) | 144 N-m |
| Rotational Velocity | 183 rad/sec |

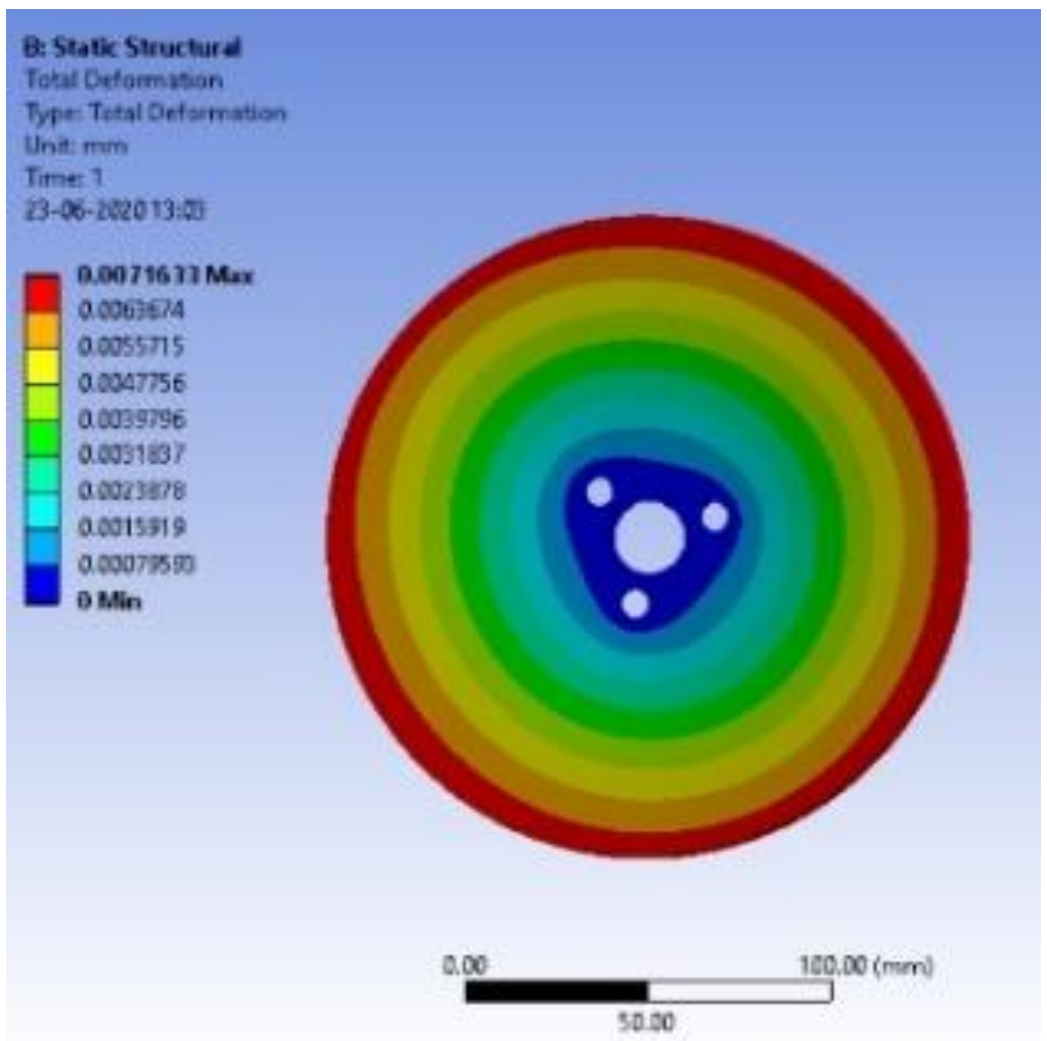

Figure 4. Deformation

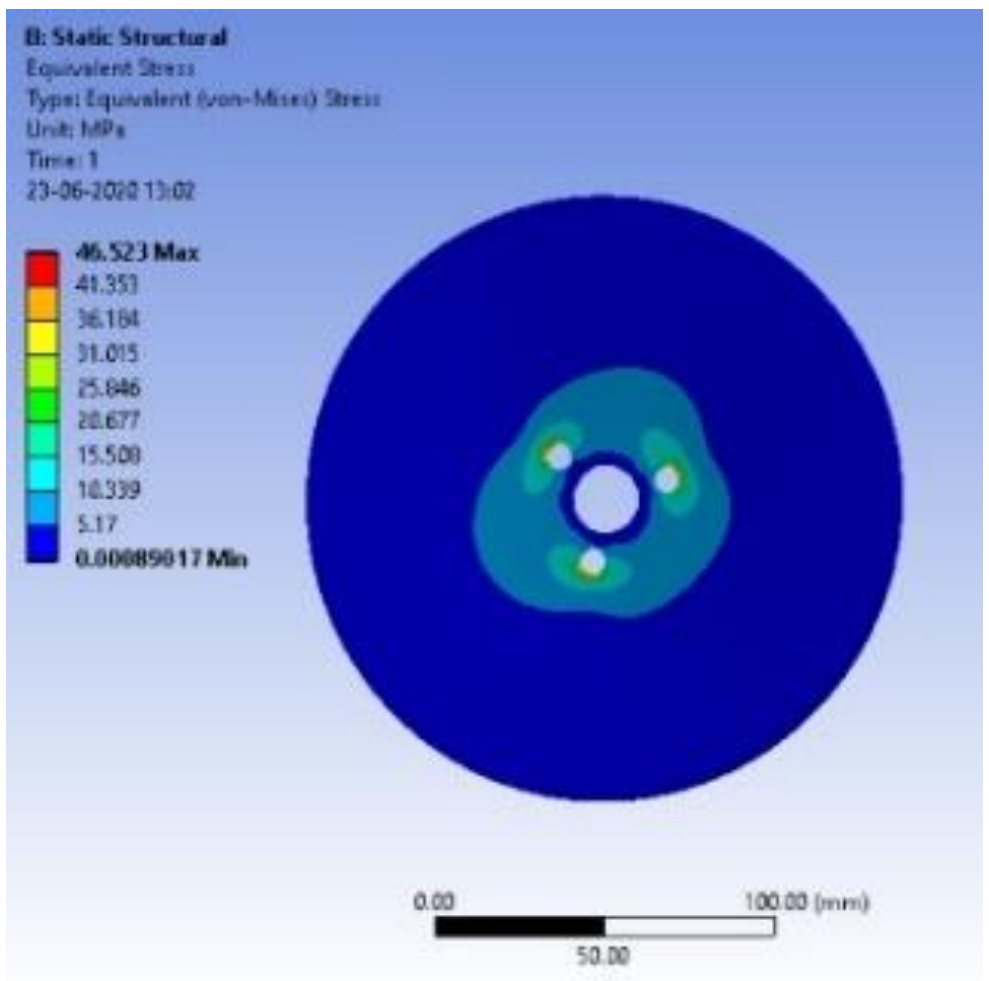

Figure 5. Equivalent Stress

Minimum factor of safety obtained from the simulation is 3.86

Structural analysis has been carried out for initial designed disc brake. Maximum equivalent stress, minimum factor of safety and total deformation contour and values are obtained from the simulation. We are able to depict the rear disc brake behaviour in the above applied conditions.

## 2.5 Thermal Analysis

**Case1: When convection is applied with the air film coefficient of 5W/m2K.**

Table 8: Thermal Brake parameters

| S.No | Parameter | Value |
|---|---|---|
| 1 | K | 1 |
| 2 | Average braking power (Watt) | 15323 |
| 3 | Average braking power absorbed per hour by rear brakes (Watt) | 3999 |
| 4 | Deceleration of Vehicle (m/sec^2) | 6.57 |
| 5 | Velocity of vehicle (m/sec) | 16 |
| 6 | Stopping time (sec) | 2.43 |
| 7 | Heat Flux (W/m2) | 68575 |

MATLAB code to obtain the heat flux using Vehicle parameters.

m= mass of vehicle; in Kg

v= velocity of vehicle; in m/sec

a= deceleration of vehicle; m/sec^ (2)

D= Outer Diameter; in m

d= Inner Diameter; in m

cr= heat lost (heat lost coefficient);

q= average Braking power; in Watt

s= Tire slip;

K=1;

q = (K*(1-s) *a*v*m)/2;

m= 310;

v= 16;

a= 6.573;

s= 0.06;

xr = weight distribution on rear wheels;

q = 1.5323e+04

qr =average braking power absorbed per hour by rear brake

qr = q*xr*cr*0.5;

xr= 0.6;

cr = 0.87;(13% Heat Lost)

qr = 3.999e+03

t= Stopping Time;

t = v/a;

t = 2.43

Final Braking Power, P1 = q1. /t;

P1 = 1645.8

Area = pie. /2 * (Dr^ (2) – dr^ (2));

Dr = 175;

dr = 20;

Area = 0.024 m^2

Heat Flux = P1/Area

Heat flux = 6.8575e+04

Table 9. Thermal simulation parameters (Normal Convection)

| Parameter | Value |
|---|---|
| Heat Flux | 68575 W/m^2 |
| Radiation to atmosphere | 22 °C |

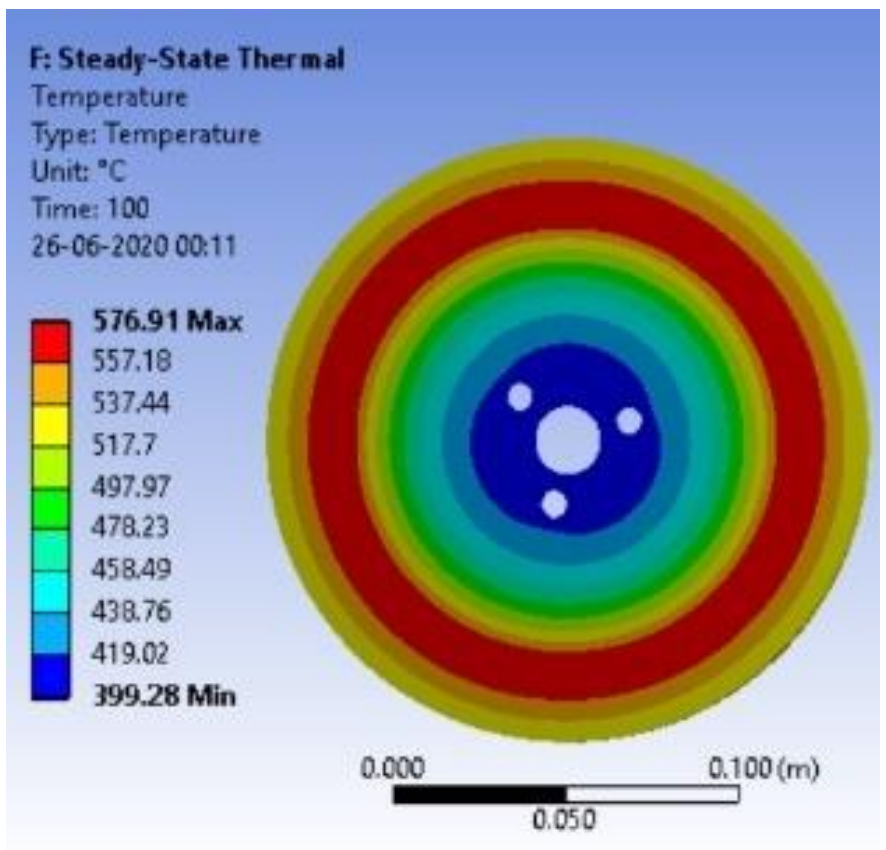

Figure 6. Temperature Contour

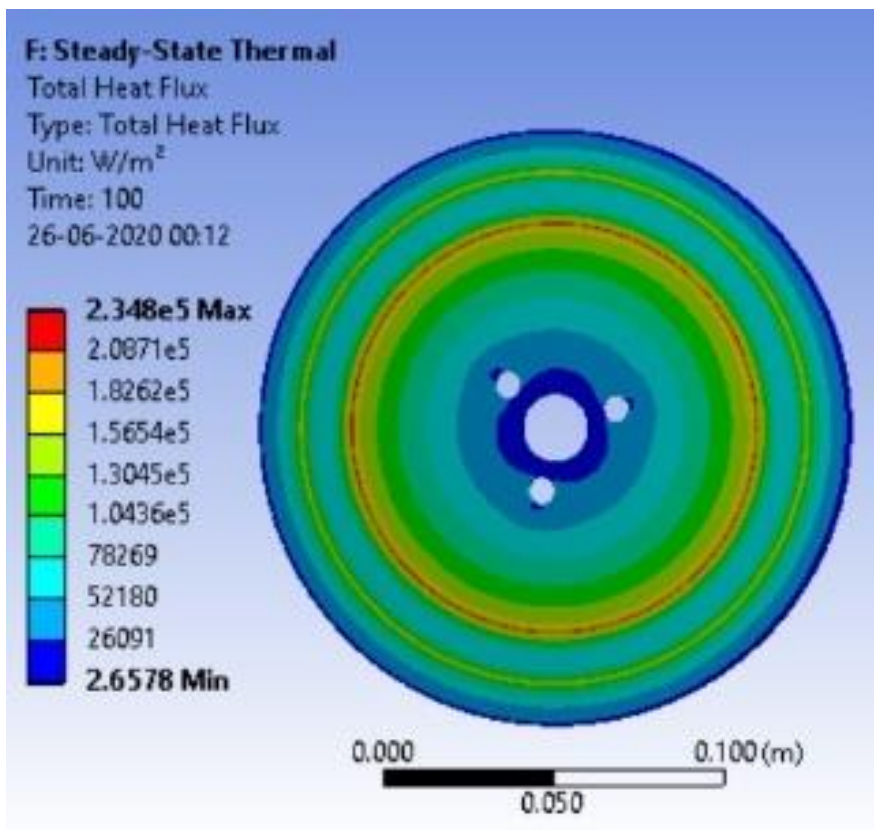

Figure 7. Total Heat Flux Contour

Thermal analysis is conducted to study the behaviour of initially designed rear disc brake. Since, laminar flow has been considered for rear disc brake but its behaviour is observed in case of normal convection and forced convection.

**Case2: Heat transfer coefficient for disc brake having laminar heat flow.**

Heat flux calculation has been shown in case 1

Convective heat transfer coefficient

Heat transfer coefficient for disc brake having laminar heat flow (Re> 2.4×10e5)

Hr =0.04(ka/D) Re^ (0.8)

Hr = 322.834 W/m^2.K

Table 10. Thermal simulation parameters (Laminar Convection)

| Parameter | Value |
|---|---|
| Heat Flux | 68575 W/m^2 |
| Radiation to atmosphere | 22 °C |
| Convection | 322.834 W/m^2.K |

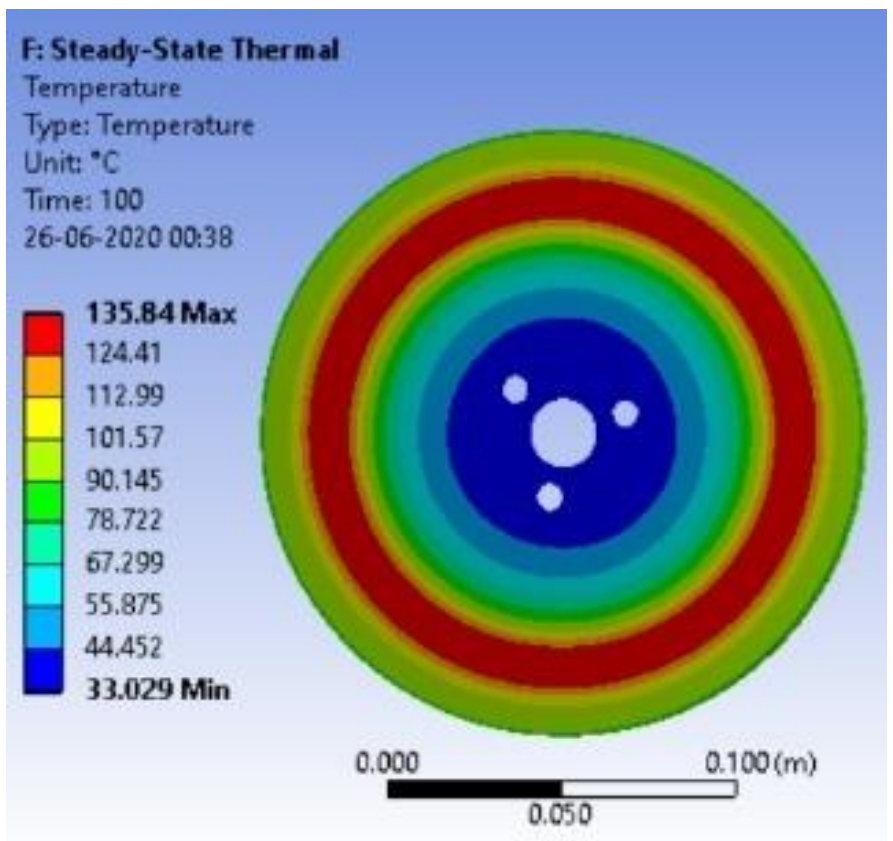

Figure 8. Temperature contour

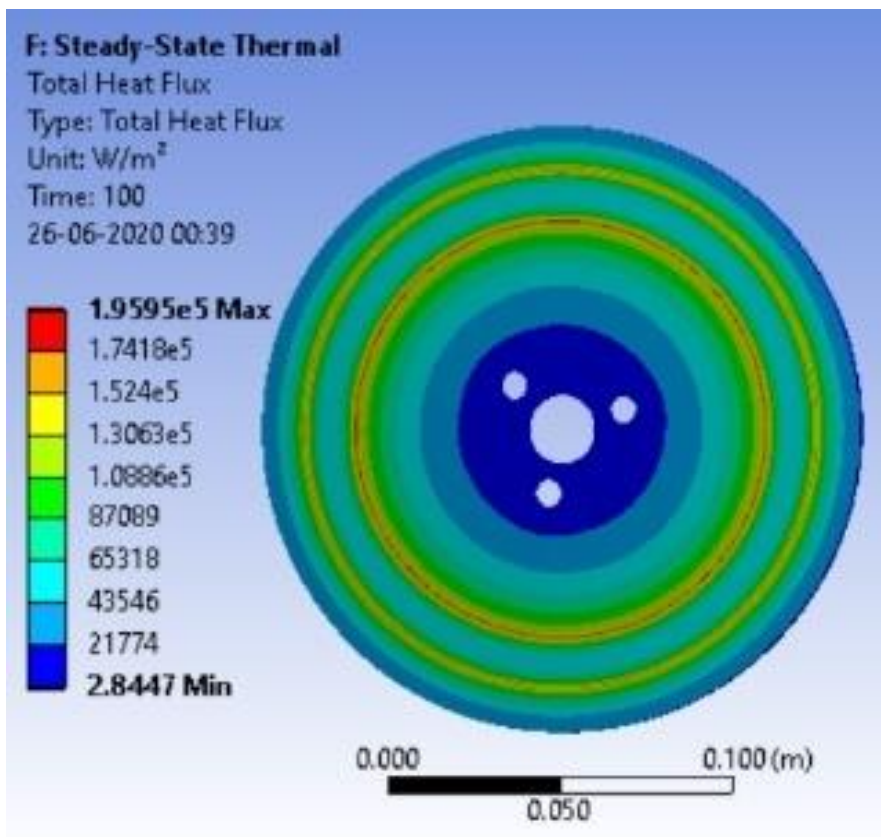

Figure 9. Total Heat Flux contour

Thermal analysis is conducted to study the behaviour of initially designed rear disc brake. Minimum temperature comparison has been carried out for initial designed and final designed rear disc brake. Heat flux is indicating the overall distribution of heat flux across the disc brake. Optimized design parameter value (thickness and outer diameter) are obtained by DOE.

**Case3: Forced convective heat transfer coefficient for disc brake.**

Heat flux calculation has been shown in case 1.

We have considered warping temperature of SS-410 to calculate the film temperature, assuming the ambient or surrounding temperature to be 300 K. Warping temperature is the temperature at which deformation just begins and it is generally numerically equal to 70% of the melting temperature of the metal.

Tmelt for SS-410 = 1500°C.

Twarp for SS-410 = 70% (1538)

Twarp for SS-410 = 1050°C. = 1323 K

Ambient temperature = Tamb = 295 K.

Film temperature = (Twarp + Tamb)/2

Film temperature = 809 K

Thus, for the required calculations for the heat transfer coefficient at the film temperature, the air properties at this film temperature is 809 K.

Relative velocity of air (v) = 16 m/s

Diameter of the rotor = 0.175 m

Reynold’s Number, Re = p*v*d/u

Re = (16 x 0.438 x 0.175)/ (36.9 x 10-6)

Re = 33235.7724

Nusselt Number, Nu = 0.0266 (Re)^ 0.805 x (Pr)^ 0.333

Nu = 0.0266 (33235.7724) ^ 0.805 x (0.718) ^ 0.333

Nu = 103.959

Forced Convective Heat Transfer Coefficient, h,

h = (Nu*k)/d

h = 103.959 x 0.059835/0.175

h = 35.5452 W/m2-K

Table 11. Thermal simulation parameters (Forced Convection)

| Parameter | Value |
|---|---|
| Heat Flux | 68575 W/m2 |
| Radiation to atmosphere | 22 °C |
| Convection | 35.5452 W/m2-K |

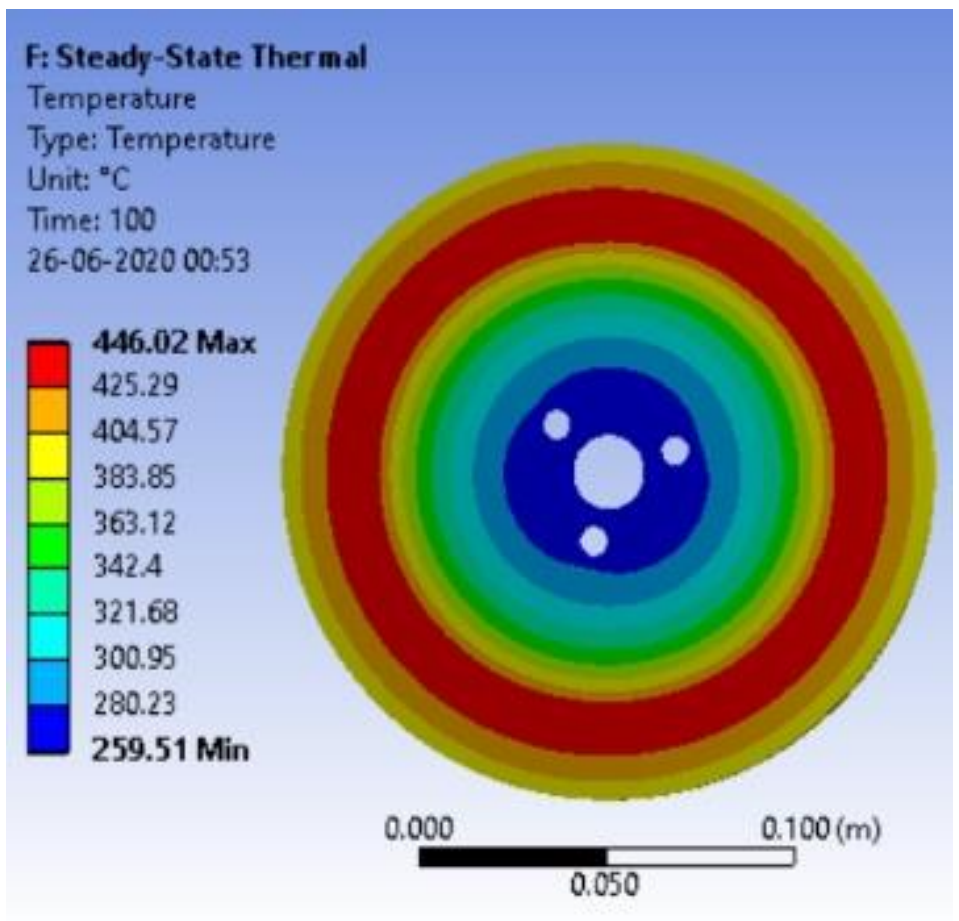

Fig10. Temperature Contour

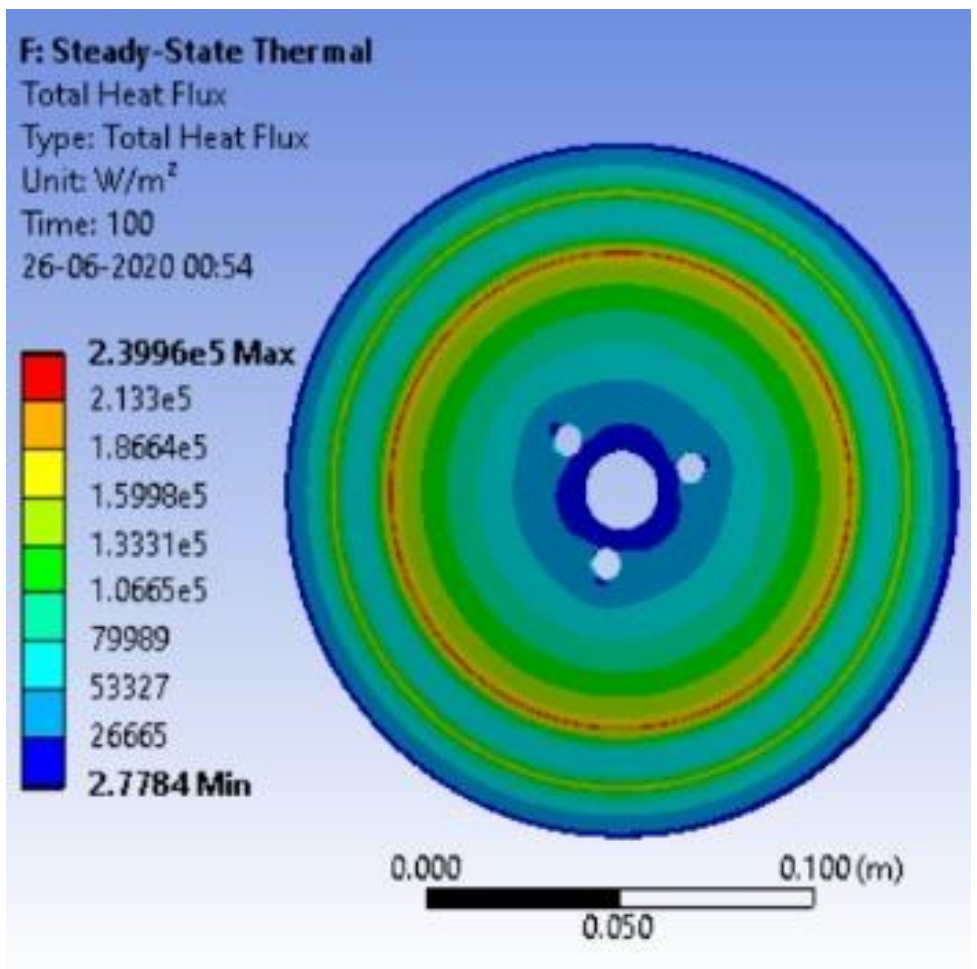

Fig11. Total Heat Flux Contour

## 2.6 Vibrational Analysis

Table 12. Frequency with respect to mode

| MODE | FREQUENCY |
|---|---|
| 1 | 701.59 Hz |
| 2 | 701.59 Hz |
| 3 | 729.52 Hz |
| 4 | 779.73 Hz |
| 5 | 958.23 Hz |
| 6 | 1548.2 Hz |

## 2.7 Goodness of fit

### 2.7.1 Goodness of fit plot for structural analysis

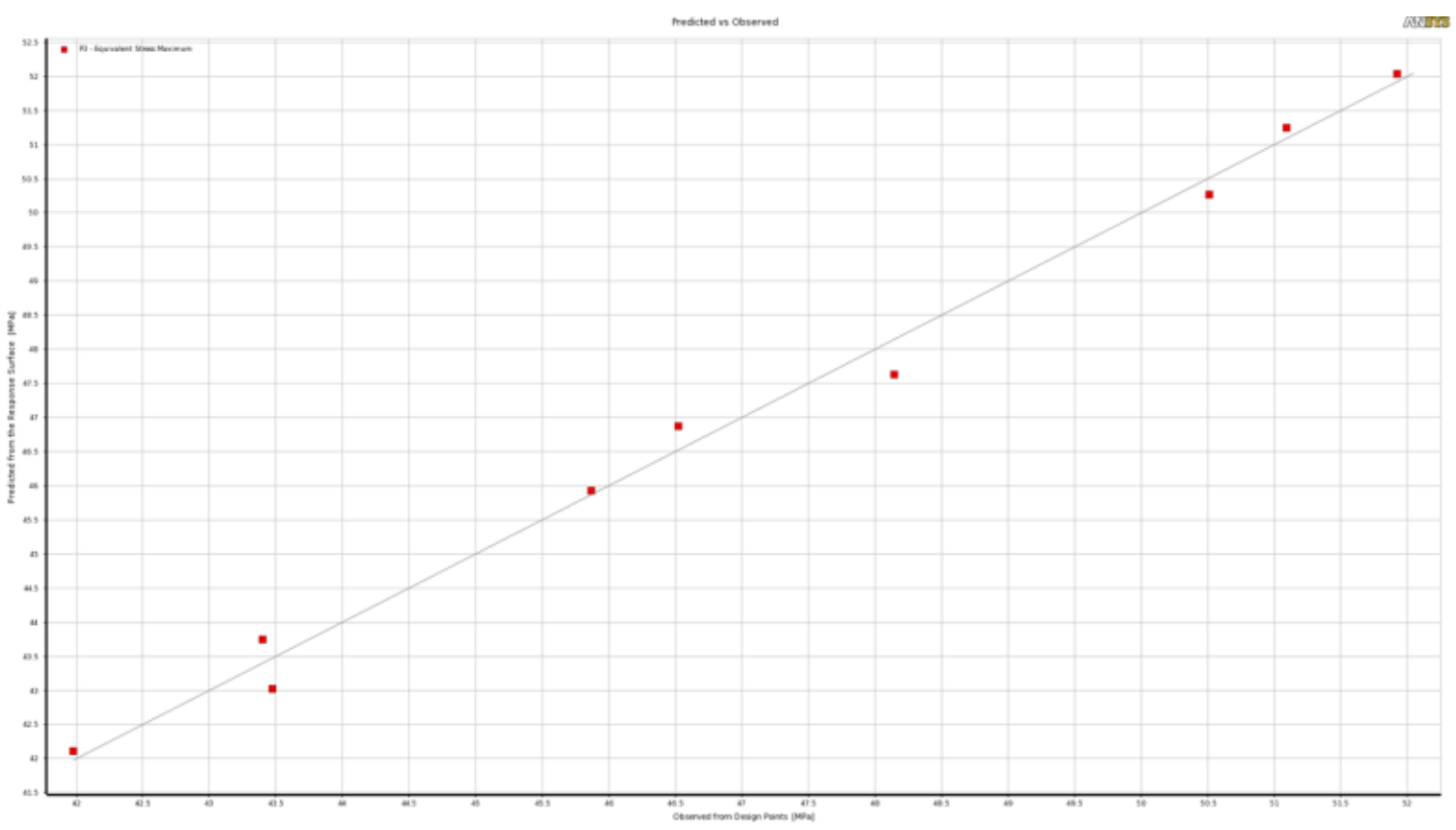

Table of Outline A12: Goodness Of Fit

| | A | B |
|---|---|---|
| 1 | Name | P3 - Equivalent Stress Maximum |
| 2 | Goodness Of Fit | |
| 3 | Coefficient of Determination (Best Value = 1) | 0.99218 |
| 4 | Adjusted Coeff of Determination (Best Value = 1) | 0.98957 |
| 5 | Maximum Relative Residual (Best Value = 0%) | 1.0645 |
| 6 | Root Mean Square Error (Best Value = 0) | 0.30459 |

Figure12. Goodness of fit

### 2.7.2 Goodness of fit plot for Thermal analysis

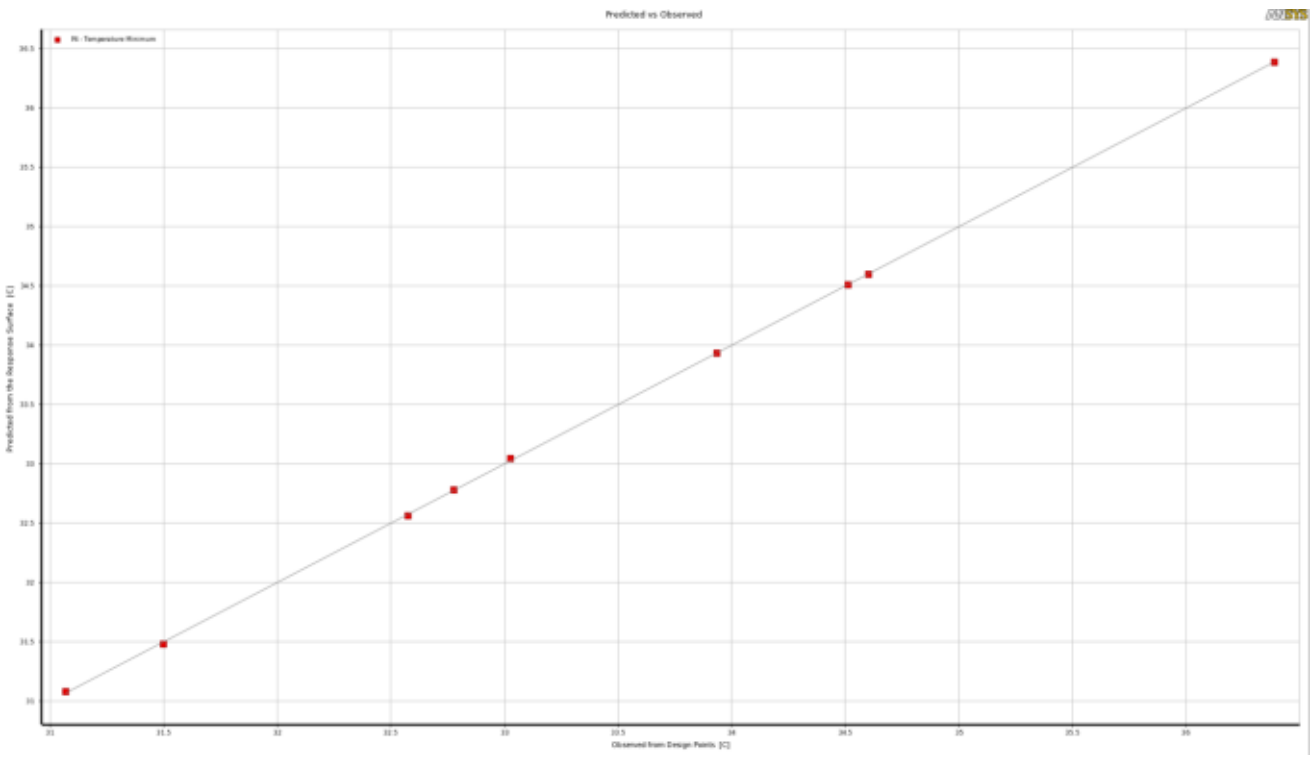

| | | |
|---|---|---|
| 2 | Goodness Of Fit | |
| 3 | Coefficient of Determination (Best Value = 1) | 0.99996 |
| 4 | Adjusted Coeff of Determination (Best Value = 1) | 0.99992 |
| 5 | Maximum Relative Residual (Best Value = 0%) | 0.04999 |
| 6 | Root Mean Square Error (Best Value = 0) | 0.0097414 |

Figure13. Goodness of fit

### 2.7.3 Goodness of fit plot for modal analysis

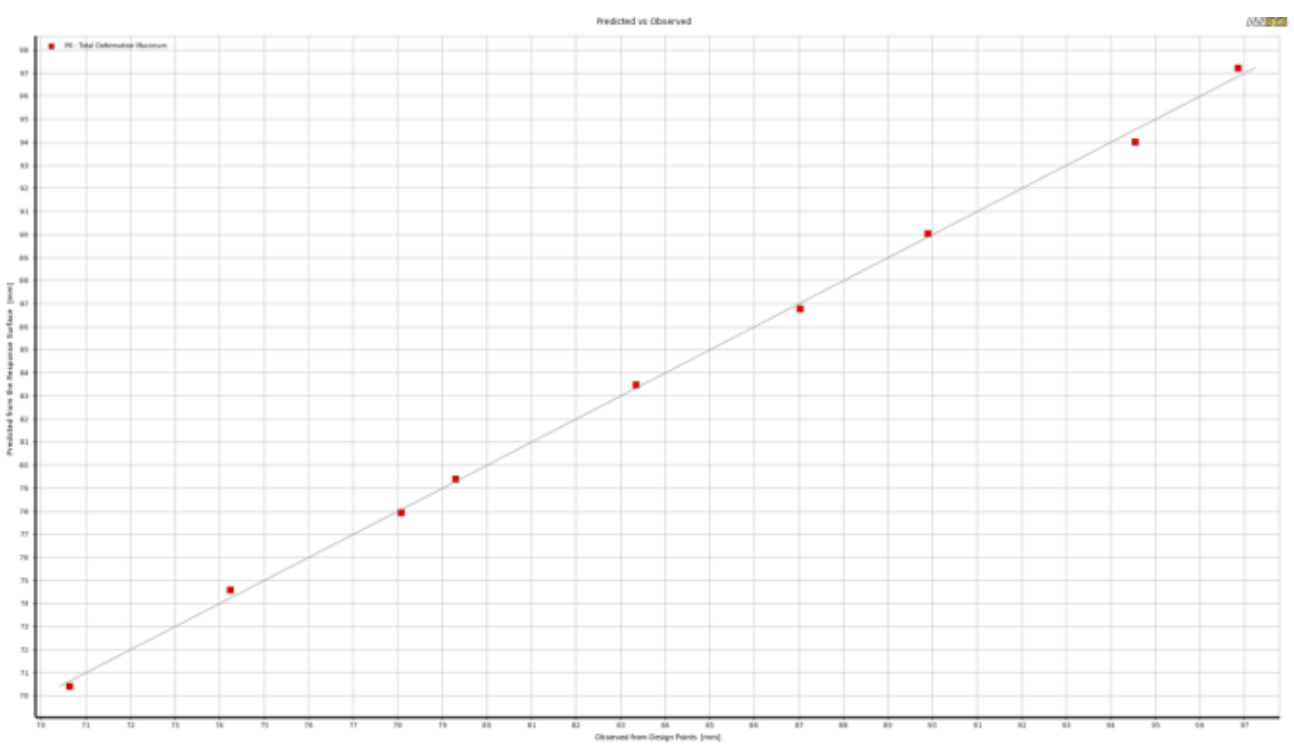

Figure14. Goodness of fit

## 2.8 Design Optimization

### 2.8.1 Structural analysis optimization

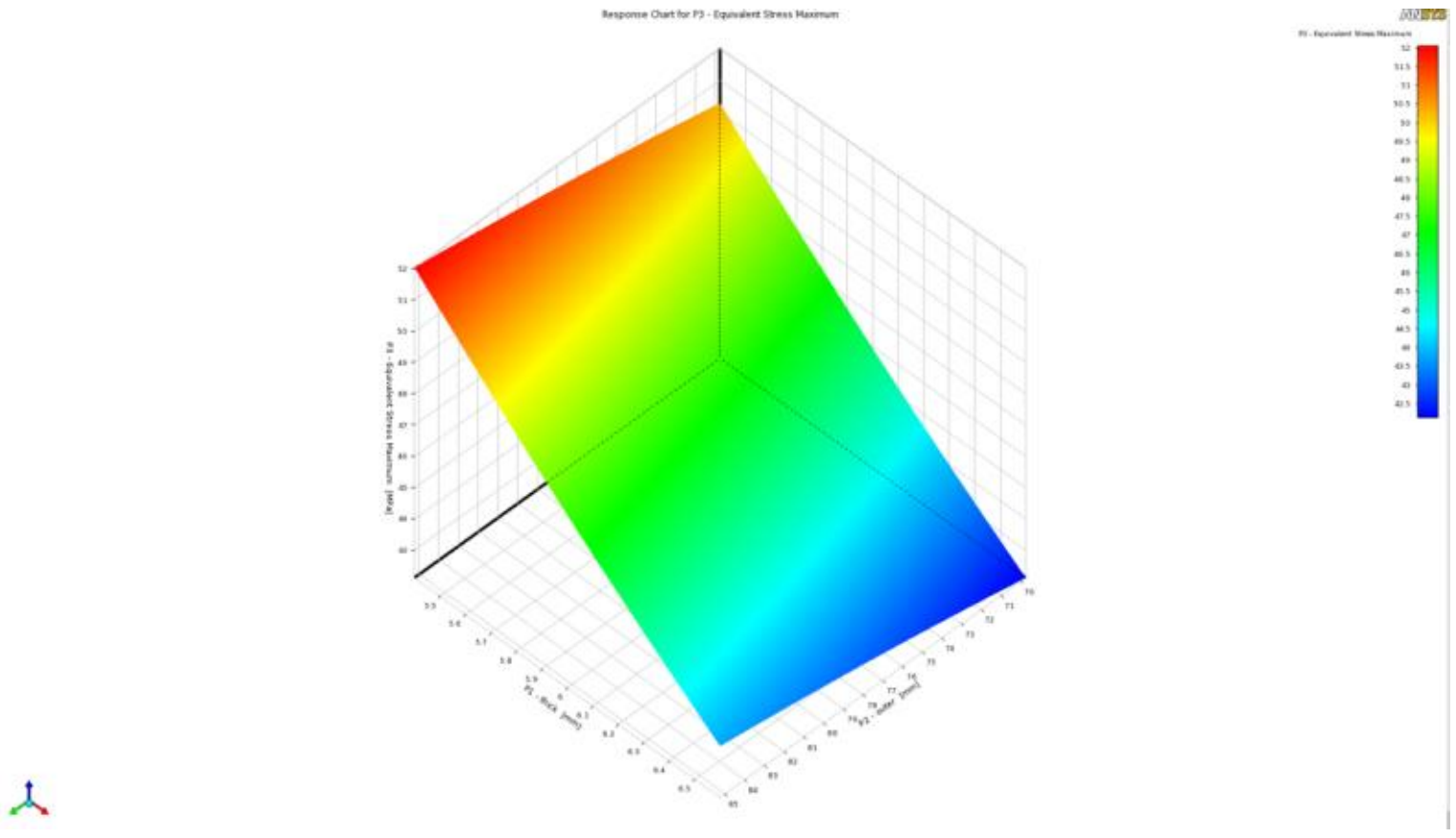

Figure 15. 3-D Plot (Equivalent Stress)

Table of Schematic H2: Design of Experiments (Central Composite Design : Auto Defined)

| | A | | B | C | D |
|---|---|---|---|---|---|
| 1 | Name | | P1 - thick (mm) | P2 - outer (mm) | P3 - Equivalent Stress Maximum (MPa) |
| 2 | 1 | DP 0 | 6 | 77.5 | 46.523 |
| 3 | 2 | DP 2 | 5.4 | 77.5 | 51.093 |
| 4 | 3 | DP 7 | 6.6 | 77.5 | 43.475 |
| 5 | 4 | DP 4 | 6 | 69.75 | 45.872 |
| 6 | 5 | DP 5 | 6 | 85.25 | 48.147 |
| 7 | 6 | DP 1 | 5.4 | 69.75 | 50.512 |
| 8 | 7 | DP 6 | 6.6 | 69.75 | 41.979 |
| 9 | 8 | DP 3 | 5.4 | 85.25 | 51.924 |
| 10 | 9 | DP 8 | 6.6 | 85.25 | 43.403 |

Figure 16. Data Set

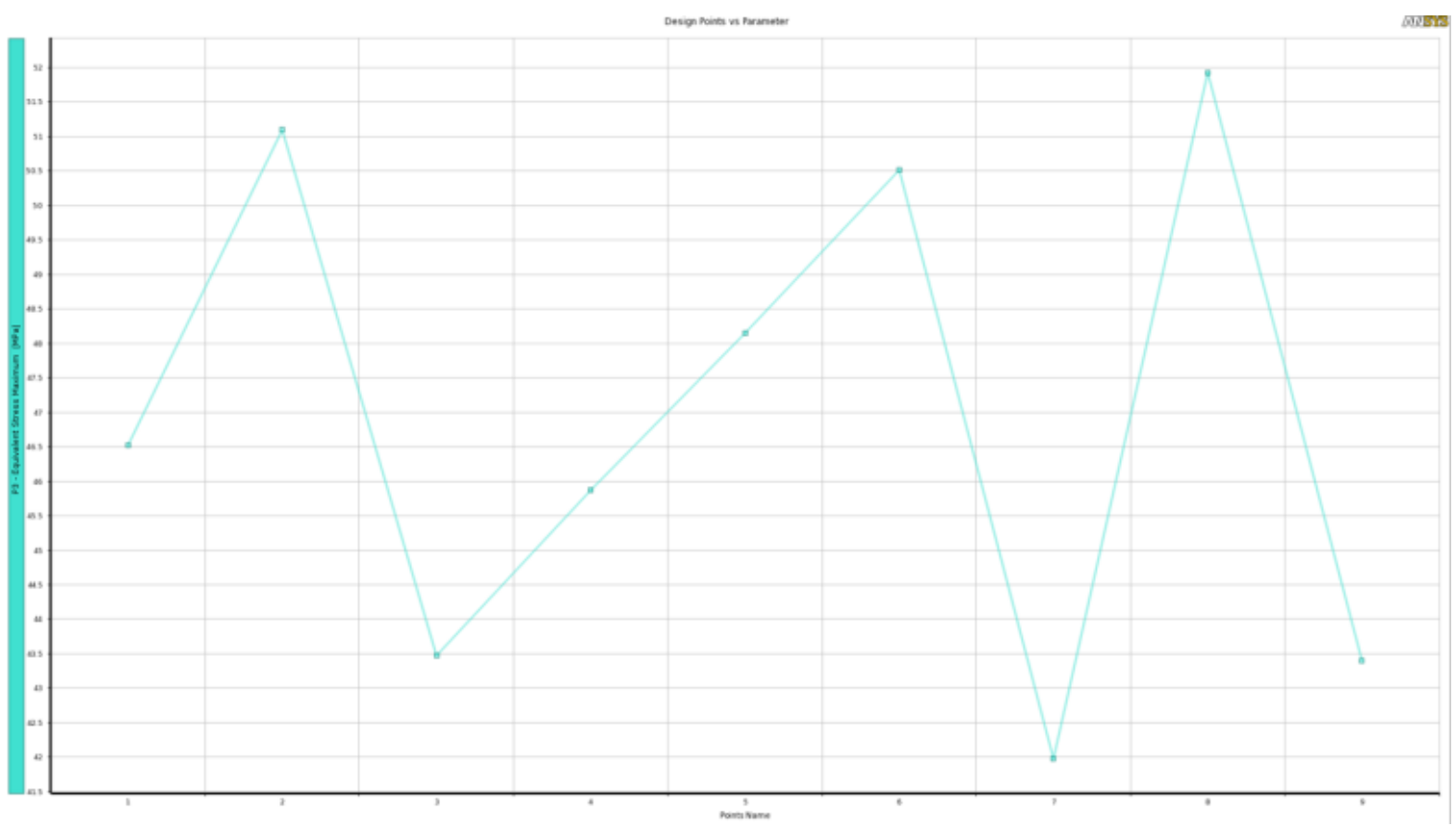

Figure17. Maximum Equivalent Stress vs Design Points

Table of Schematic I4: Optimization , Candidate Points

| | A | B | C | D | E | F |
|---|---|---|---|---|---|---|
| 1 | Reference | Name | P1 - thick (mm) | P2 - outer (mm) | P3 - Equivalent Stress Maximum (MPa) | |
| 2 | | | | | Parameter Value | Variation from Reference |
| 3 | | Candidate Point 1 | 6.6 | 69.75 | 42.118 | -1.24% |
| 4 | | Candidate Point 1 (verified) DP 6 | | | 41.979 | -1.57% |
| 5 | | Candidate Point 2 | 6.5814 | 71.438 | 42.442 | -0.48% |
| 6 | | Candidate Point 3 | 6.5862 | 73.375 | 42.648 | 0.00% |
| * | | New Custom Candidate Point | 6 | 77.5 | | |

Figure 18. Candidate points

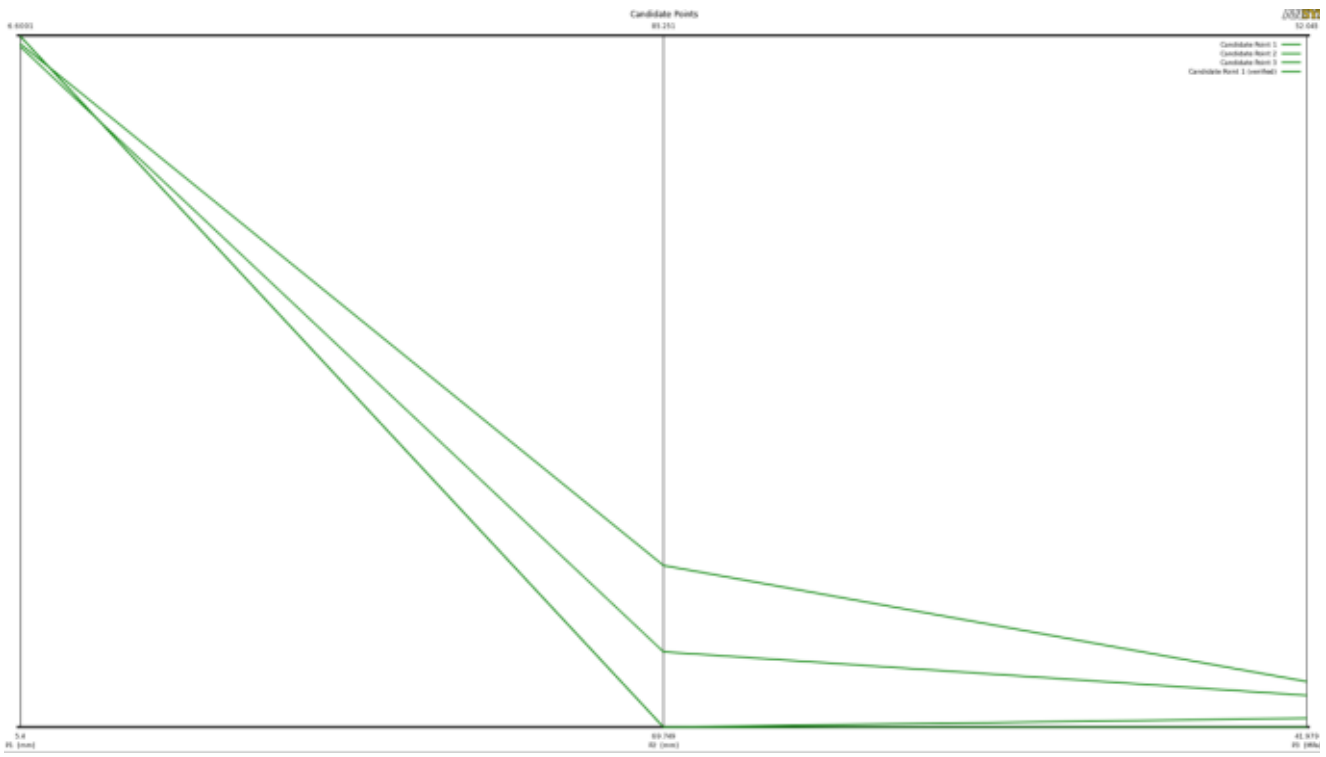

After initial analysis of each subsystem, relationship between design variables and output response is determined. All the input variables are quantitative and continuous in nature. To obtain the accurate response surface, minimum number of design points from the given sample space are required. Latin Hypercube Sampling (LHS) technique with user defined sample points is used to create the response surface.

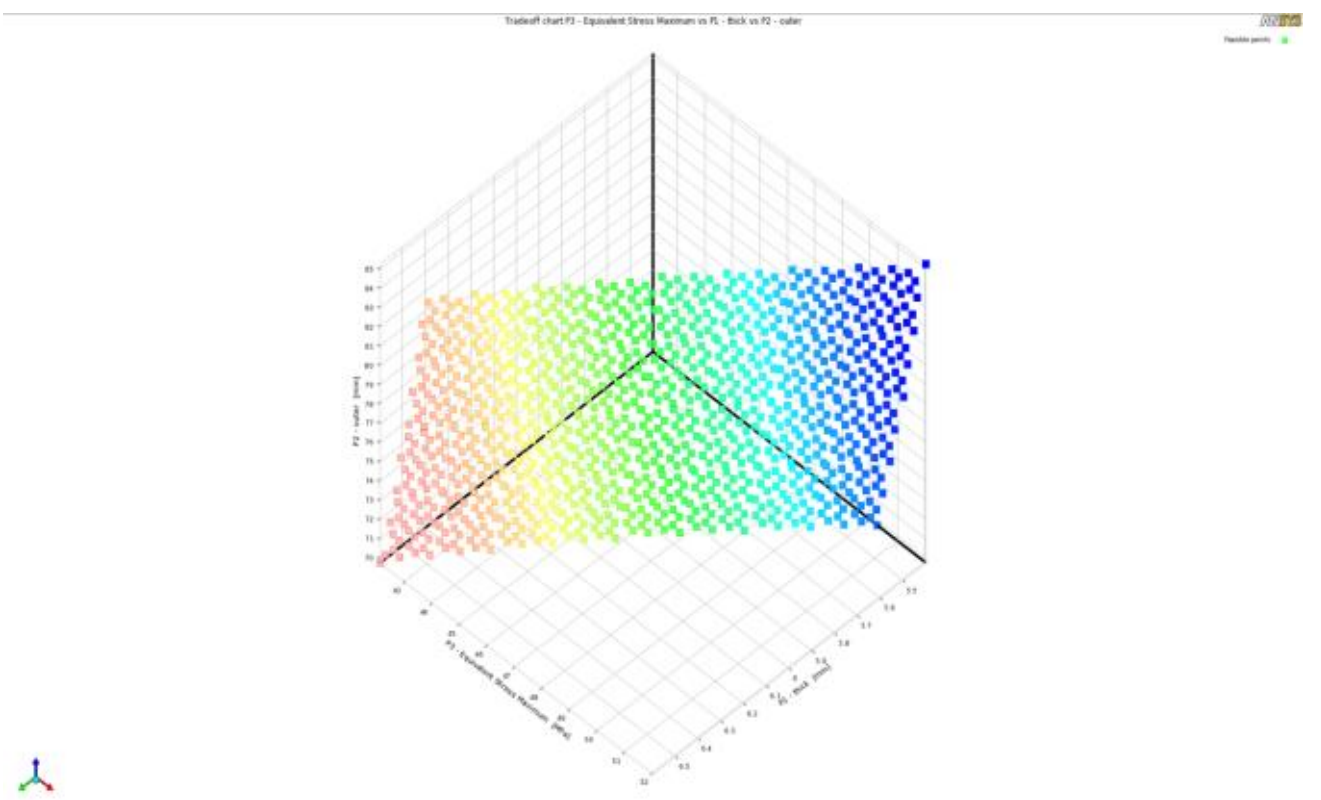

Figure 19. Trade Off Chart

2.8.2 Thermal analysis optimization

It has been carried out for Heat transfer coefficient for brake rotor having laminar flow.

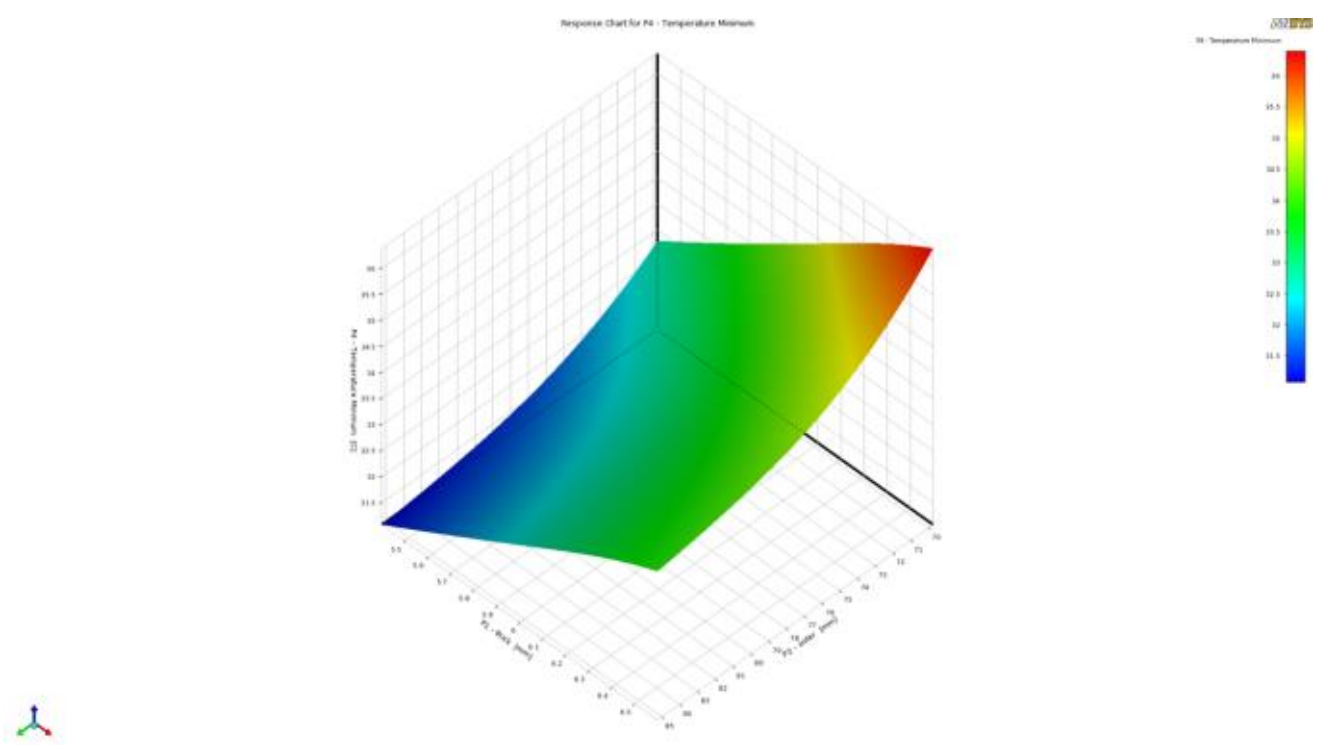

Figure 20. 3-D Plot (Minimum Temperature)

| 1 | Name | | P1 - thick (mm) | P2 - outer (mm) | P4 - Temperature Minimum (C) |
|---|---|---|---|---|---|
| 2 | 1 | DP 0 | 6 | 77.5 | 33.029 |
| 3 | 2 | DP 2 | 5.4 | 77.5 | 31.501 |
| 4 | 3 | DP 7 | 6.6 | 77.5 | 34.514 |
| 5 | 4 | DP 4 | 6 | 69.75 | 34.604 |
| 6 | 5 | DP 5 | 6 | 85.25 | 32.578 |
| 7 | 6 | DP 1 | 5.4 | 69.75 | 32.779 |
| 8 | 7 | DP 6 | 6.6 | 69.75 | 36.391 |
| 9 | 8 | DP 3 | 5.4 | 85.25 | 31.07 |
| 10 | 9 | DP 8 | 6.6 | 85.25 | 33.935 |

Figure 21. Data set

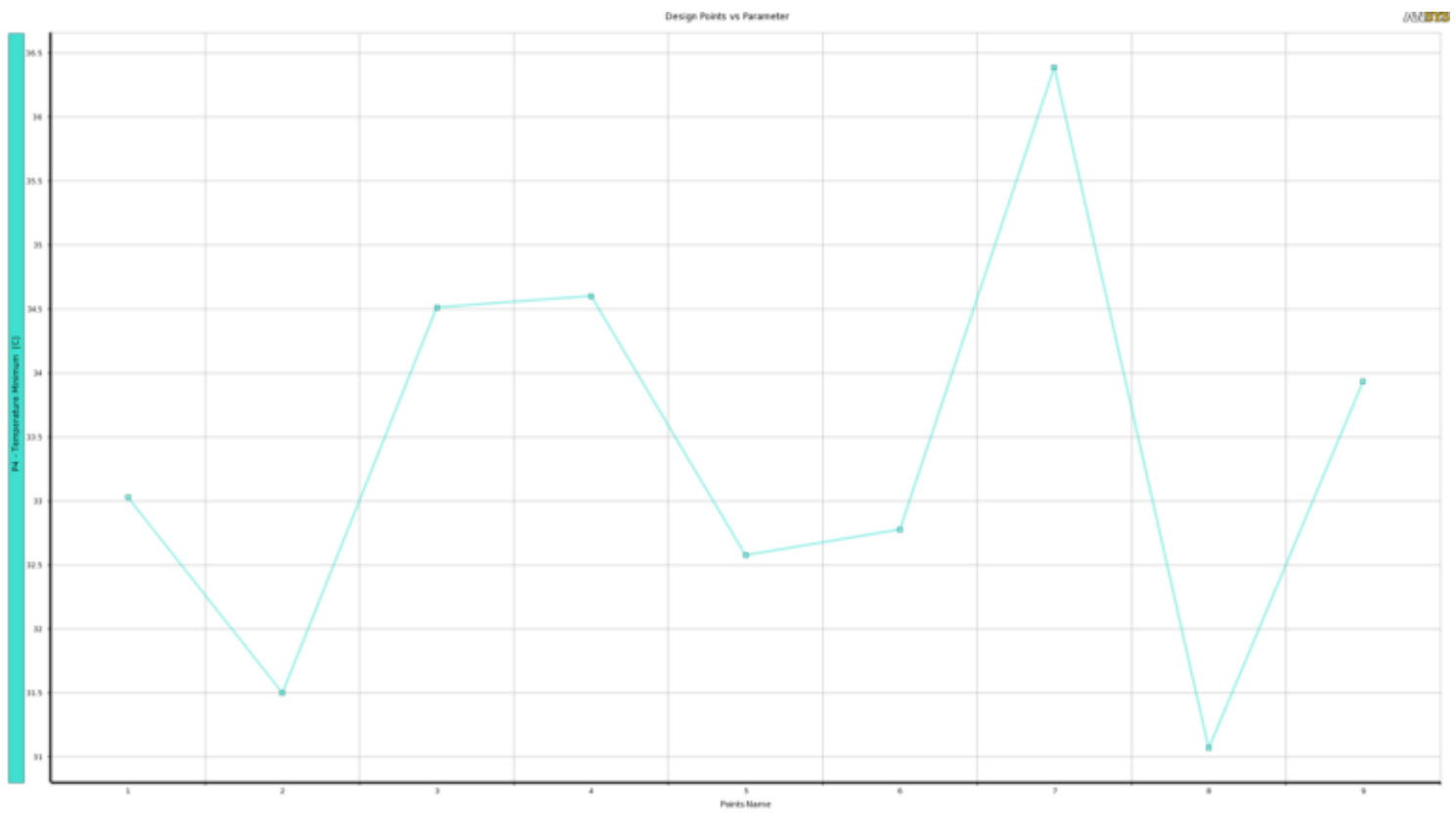

Figure22. Minimum Temperature vs Design Points

| 1 | Reference | Name | P1 - thick (mm) | P2 - outer (mm) | P4 - Temperature Minimum (C) | |
|---|---|---|---|---|---|---|
| 2 | | | | | Parameter Value | Variation from Reference |
| 3 | ○ | Candidate Point 1 | 6.045 | 78.976 | 33.028 | 0.01% |
| 4 | ○ | Candidate Point 2 | 6.0786 | 80.187 | 33.026 | 0.01% |
| 5 | ◉ | Candidate Point 3 | 5.8446 | 74.511 | 33.024 | 0.00% |
| * | | New Custom Candidate Point | 6 | 77.5 | | |

Figure 23. Candidate points

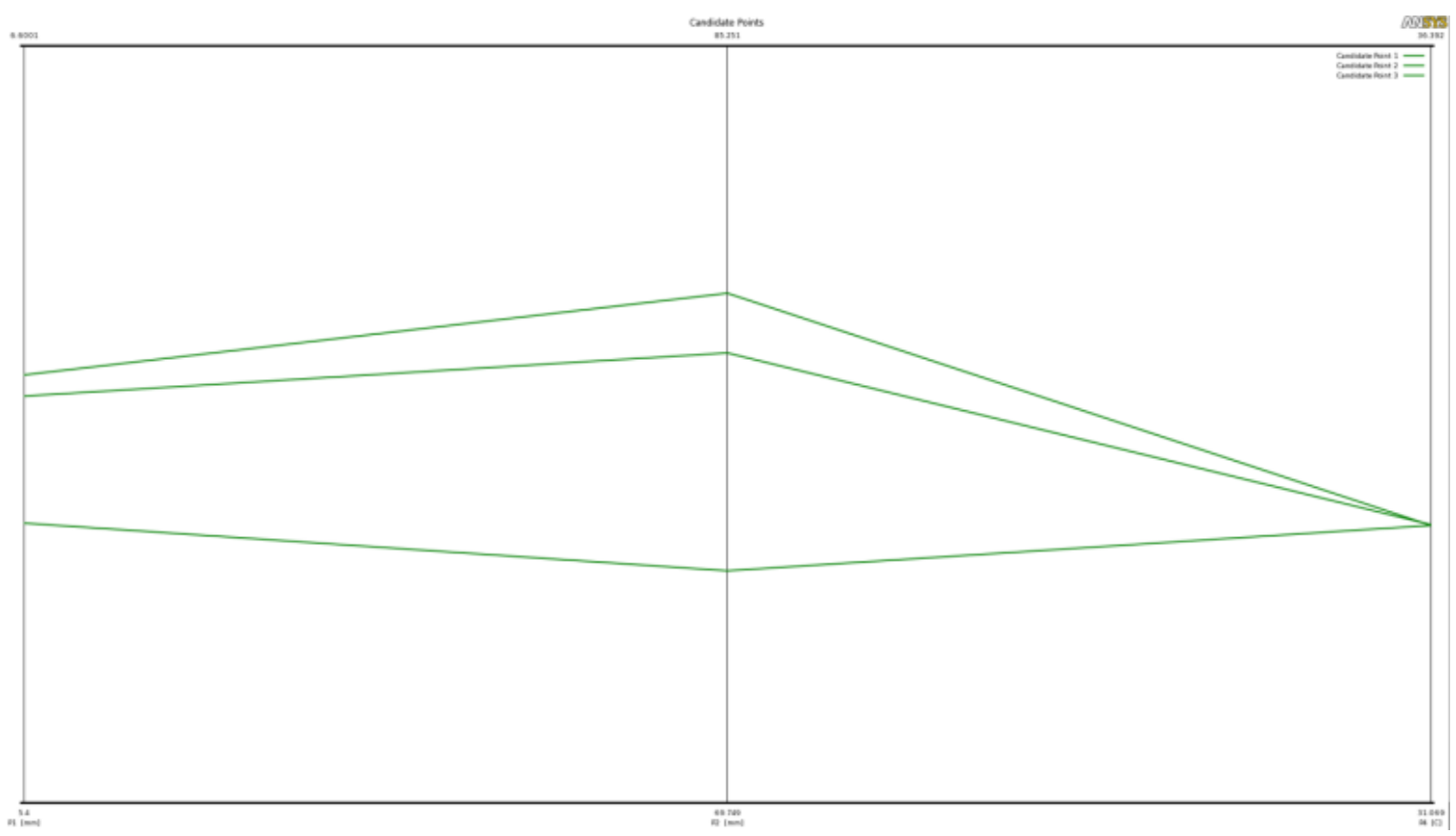

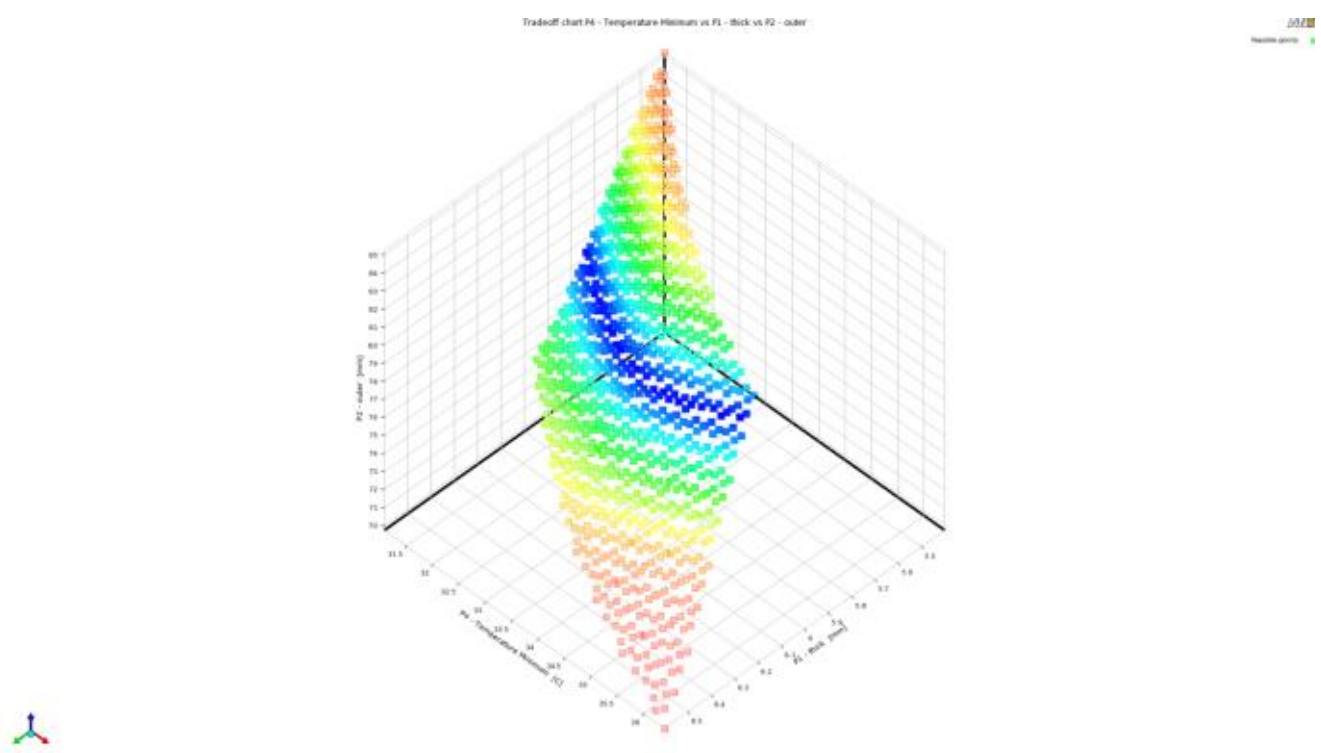

Figure 24.Trade Off Chart

2.8.3 Modal analysis optimization

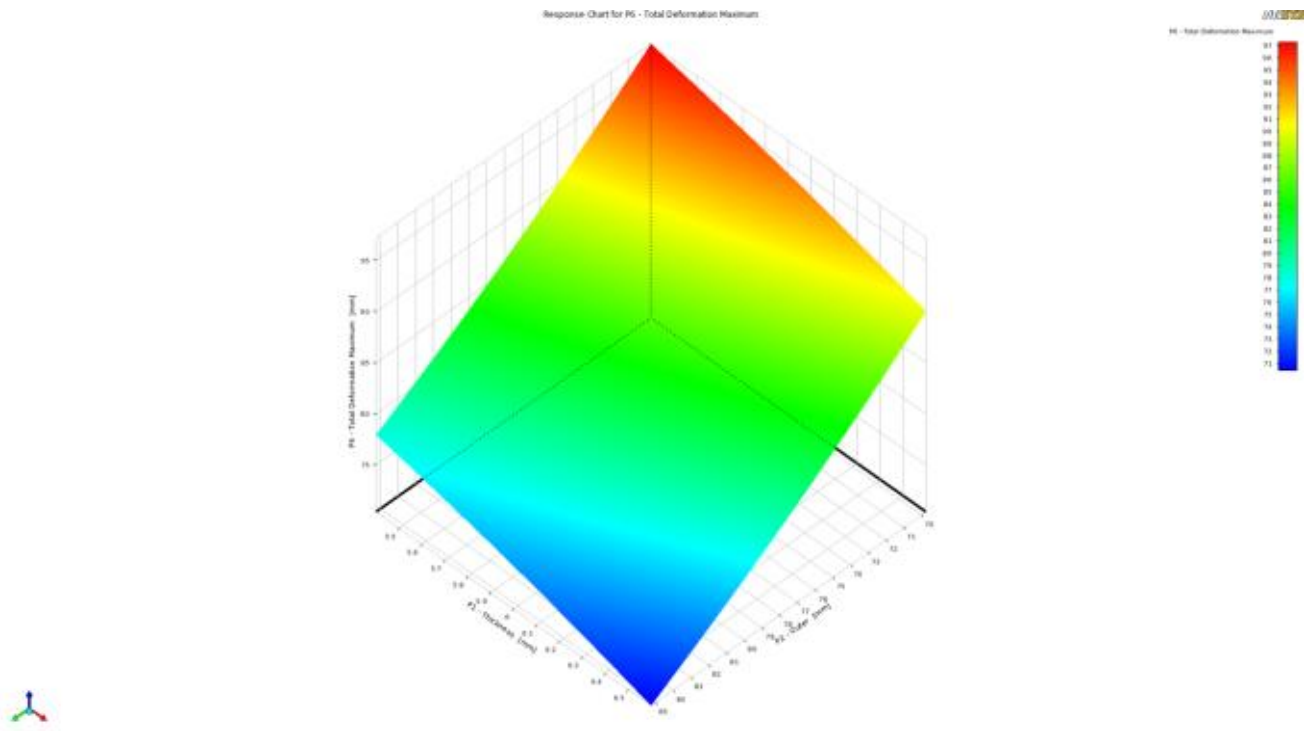

Figure 25. 3-D Plot (Maximum Total deformation 1st Order)

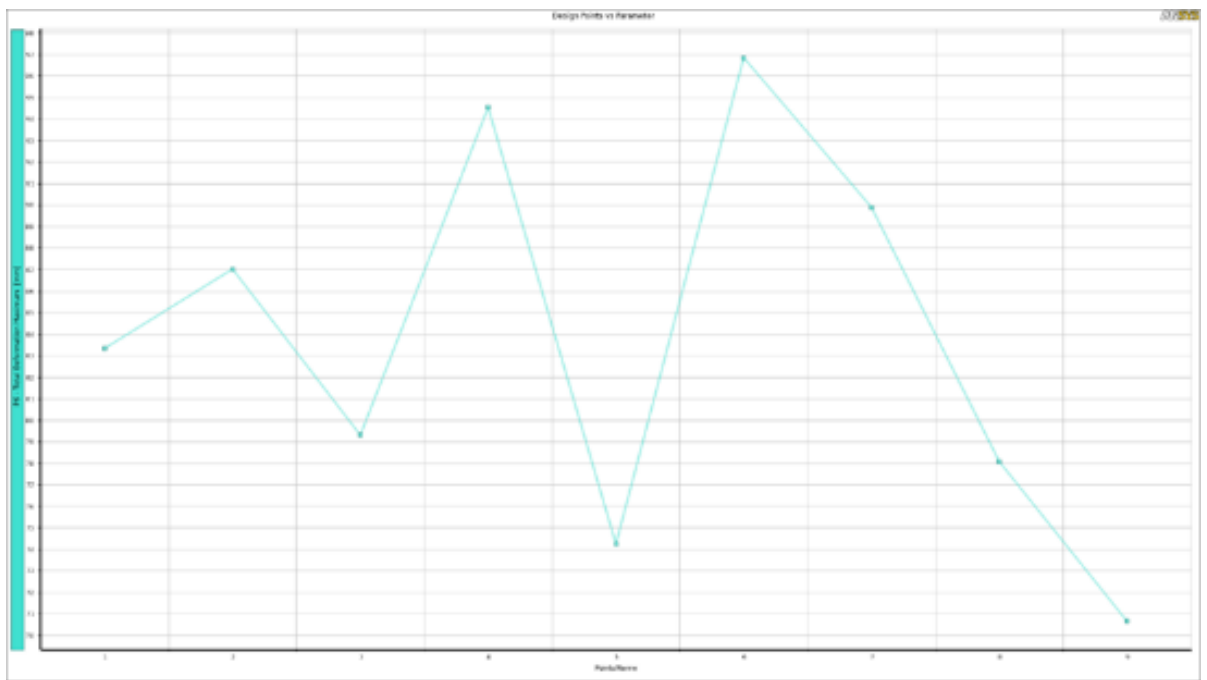

Figure 26. Total Deformation vs Design points

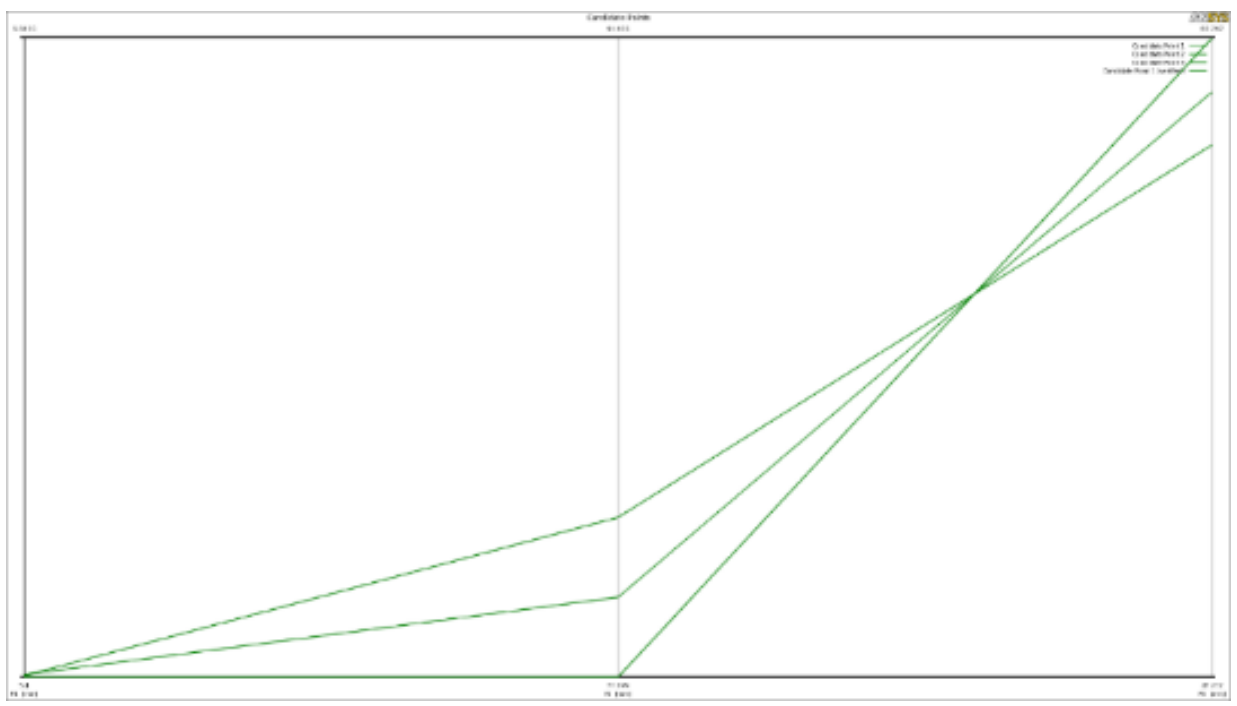

Figure 27. Candidate points

## 3. Improvized design of the brake disc

New design has been modelled after analysing all the design variables.

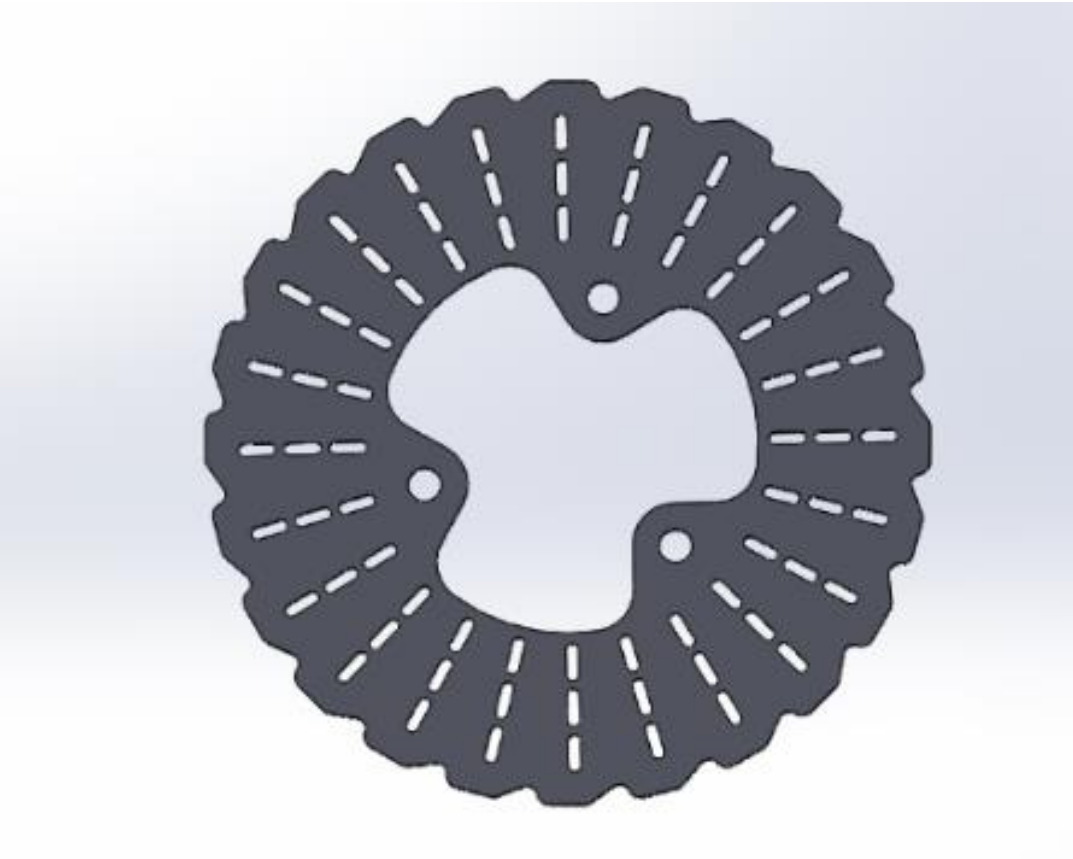

Figure 28. Improvized Design of Rear Disc Brake

Table 13. Design parameters

| S.No | Parameter | Dimension |
|---|---|---|
| 1 | Outer diameter | 175 mm |
| 2 | Inner diameter | 20 mm |
| 3 | Thickness | 5 mm |
| 4 | Front rotor (Holes size) | 8 mm |
| 5 | Number of holes | 3 |
| 6 | Number of patterns | 24 |

### 3.1 Design and Simulation

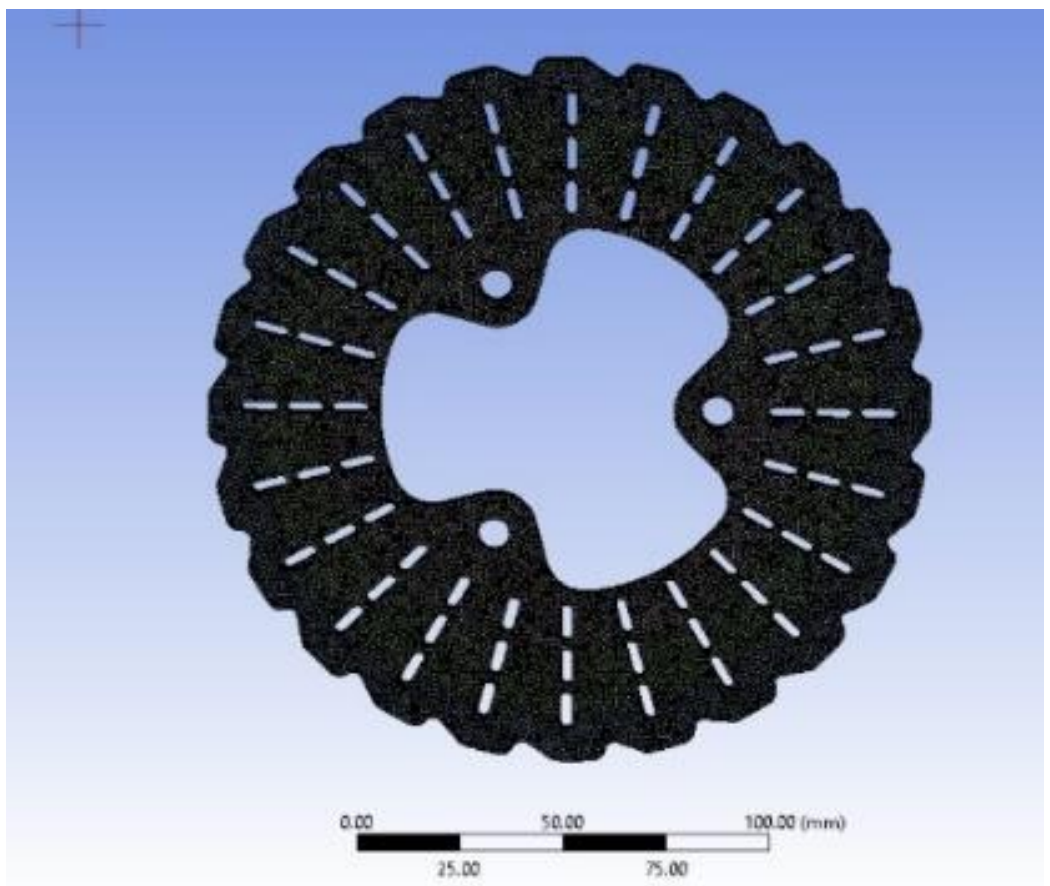

Figure 29. Meshing

Table 14. Meshing Details

| S.No | Parameter | Value |
|---|---|---|
| 1 | Nodes | 1293013 |
| 2 | Elements | 870287 |
| 3 | Method | Tetrahedron |
| 4 | Method | Body Sizing |
| 5 | Method | Fine Meshing |

## 3.2 Structural Analysis

Table 15. Structural simulation parameters

| Parameter | Value |
|---|---|
| Braking Force (Front) | 1986.2 N |
| Braking Torque (Front) | 144 N-m |
| Rotational Velocity | 183 rad/sec |

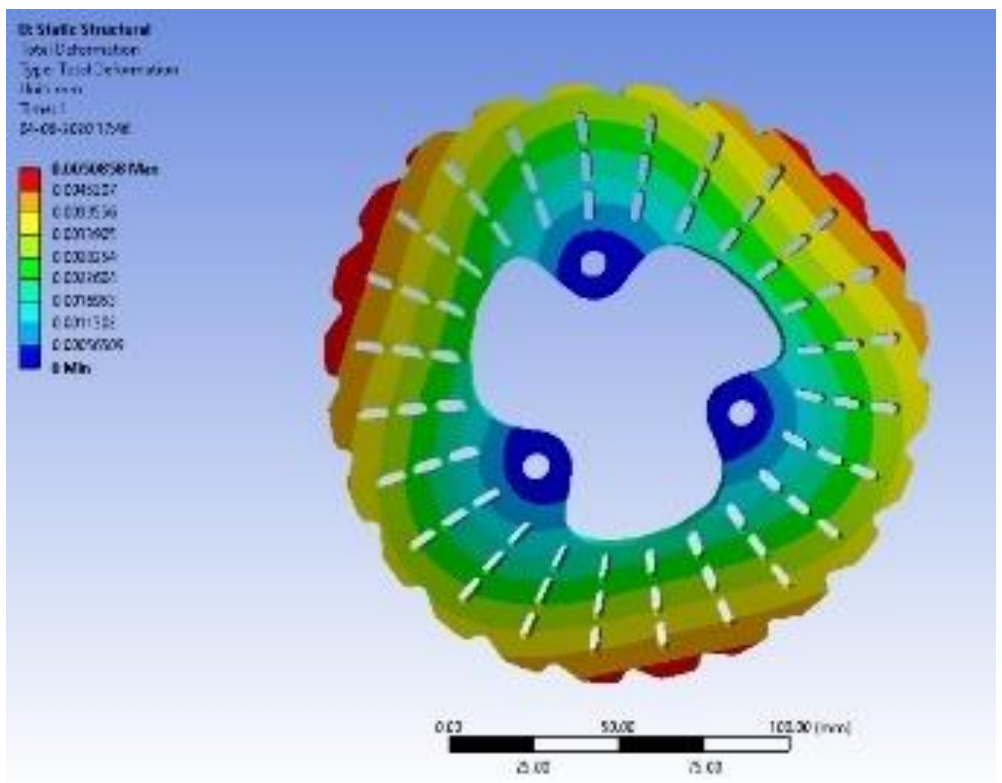

Figure 30. Total deformation

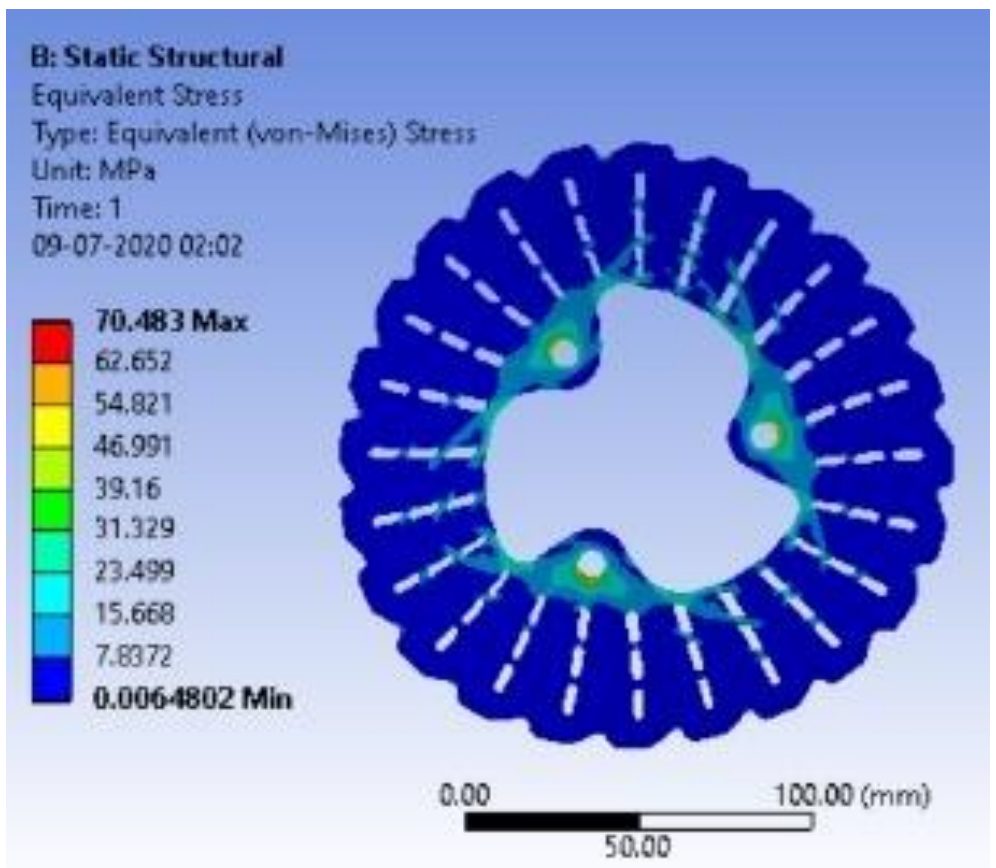

Figure 31. Equivalent Stress

Minimum factor of safety obtained from the simulation is 2.54

Structural analysis has been carried out for final designed disc brake. Maximum equivalent stress, Minimum factor of safety and total deformation contour and values are obtained from the simulation. We are able to depict the rear disc brake behaviour in above applied conditions.

## 3.3 Thermal Analysis

Heat flux and Heat Convection Calculation has been done and applied on the disc brake to obtain the simulation Output.

Case1: When convection is applied on all the surfaces with the air film coefficient of 5W/m2K

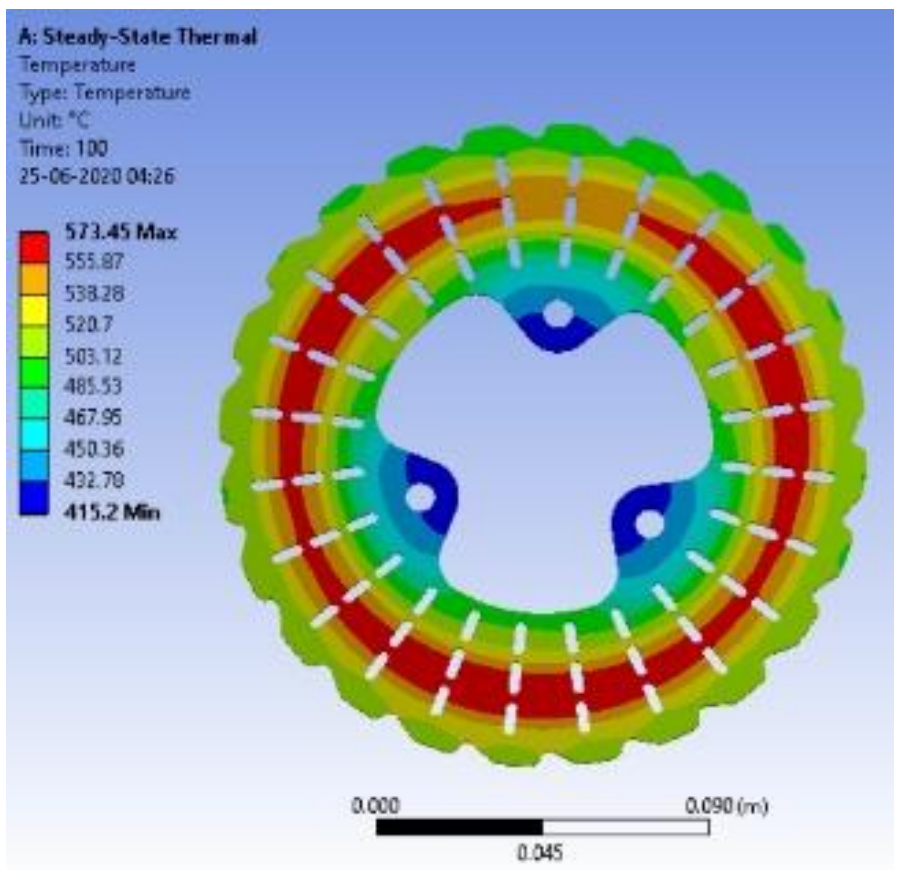

Figure 32.Temperature Contour

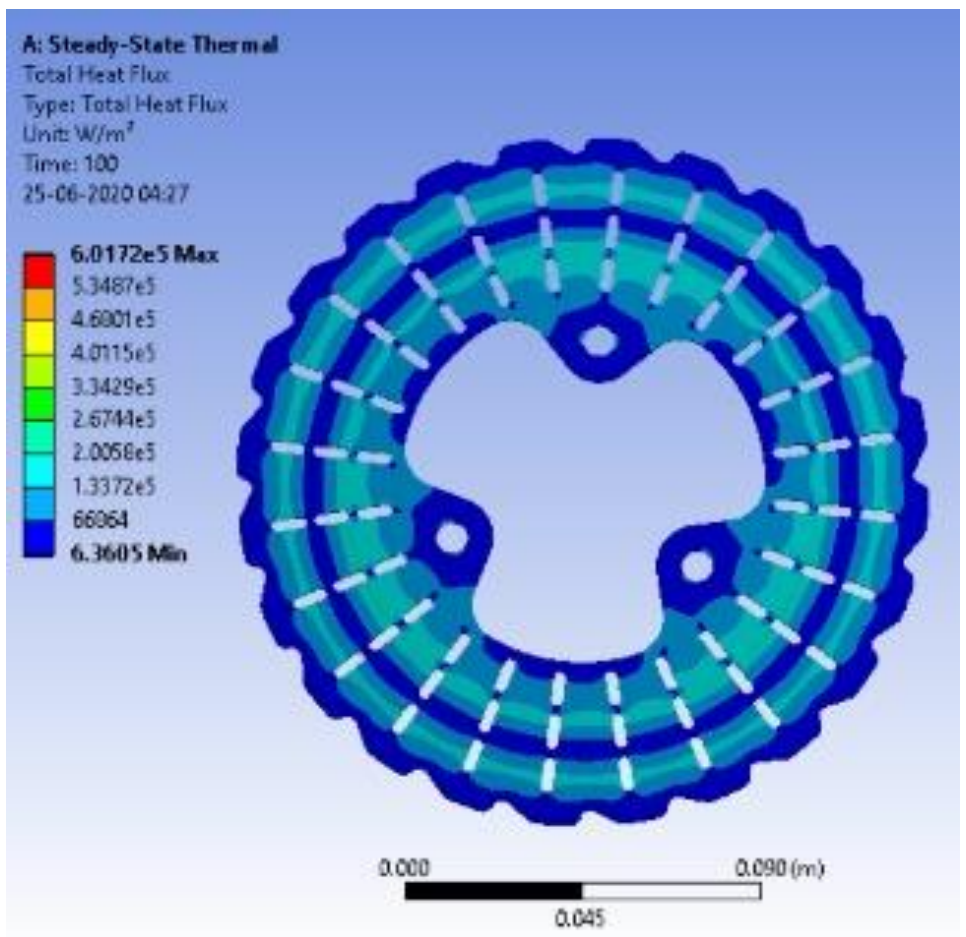

Figure 33. Heat Flux Contour

Case2: Heat transfer coefficient for disc brake having laminar heat flow**.**

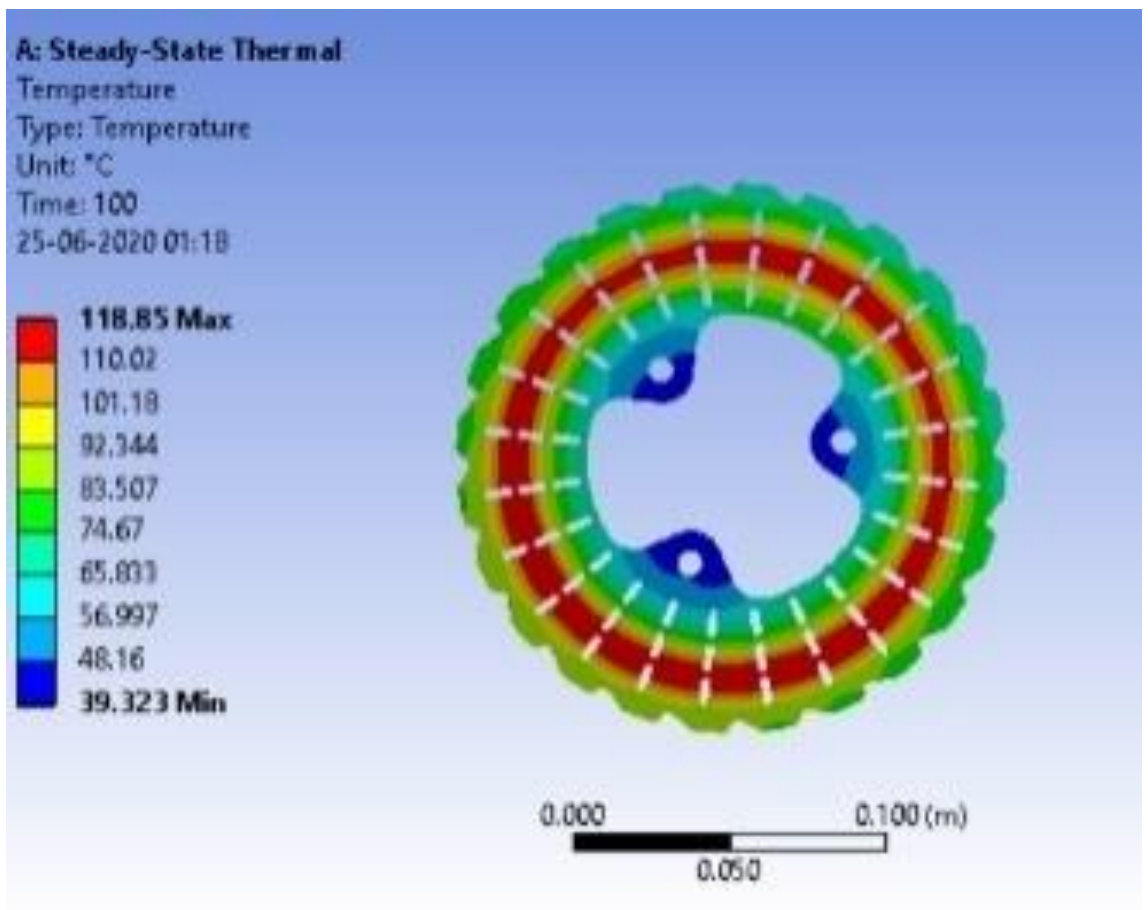

Figure 34. Temperature Contour

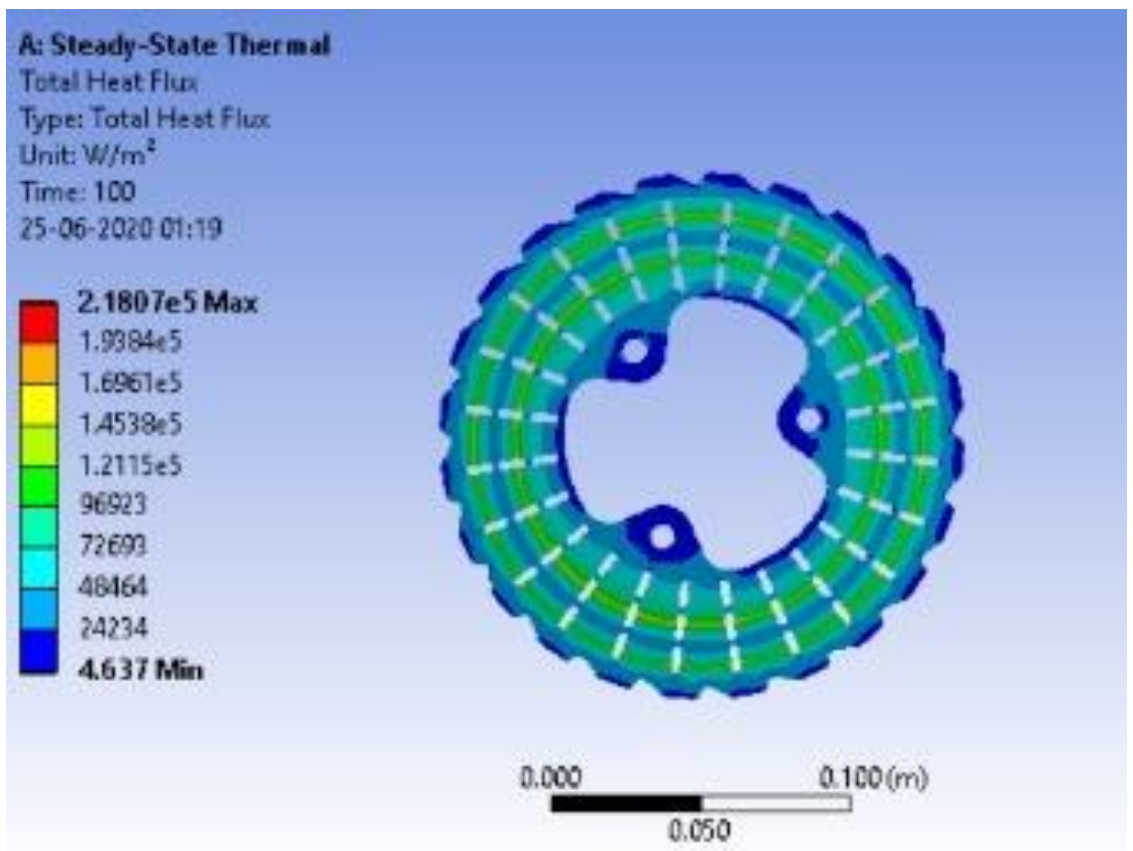

Figure 35. Total Heat Flux Contour

Case3: Forced convective heat transfer coefficient for disc brake

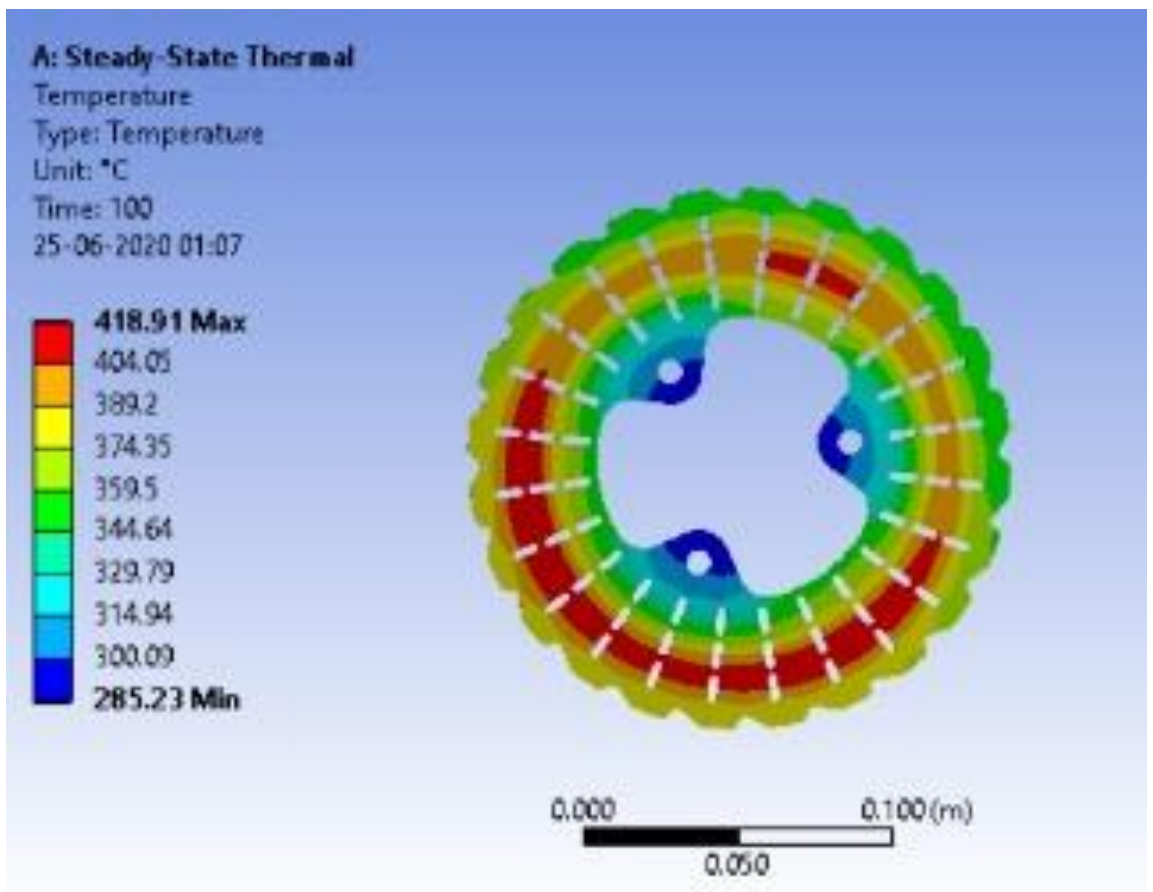

Figure 36. Temperature Contour

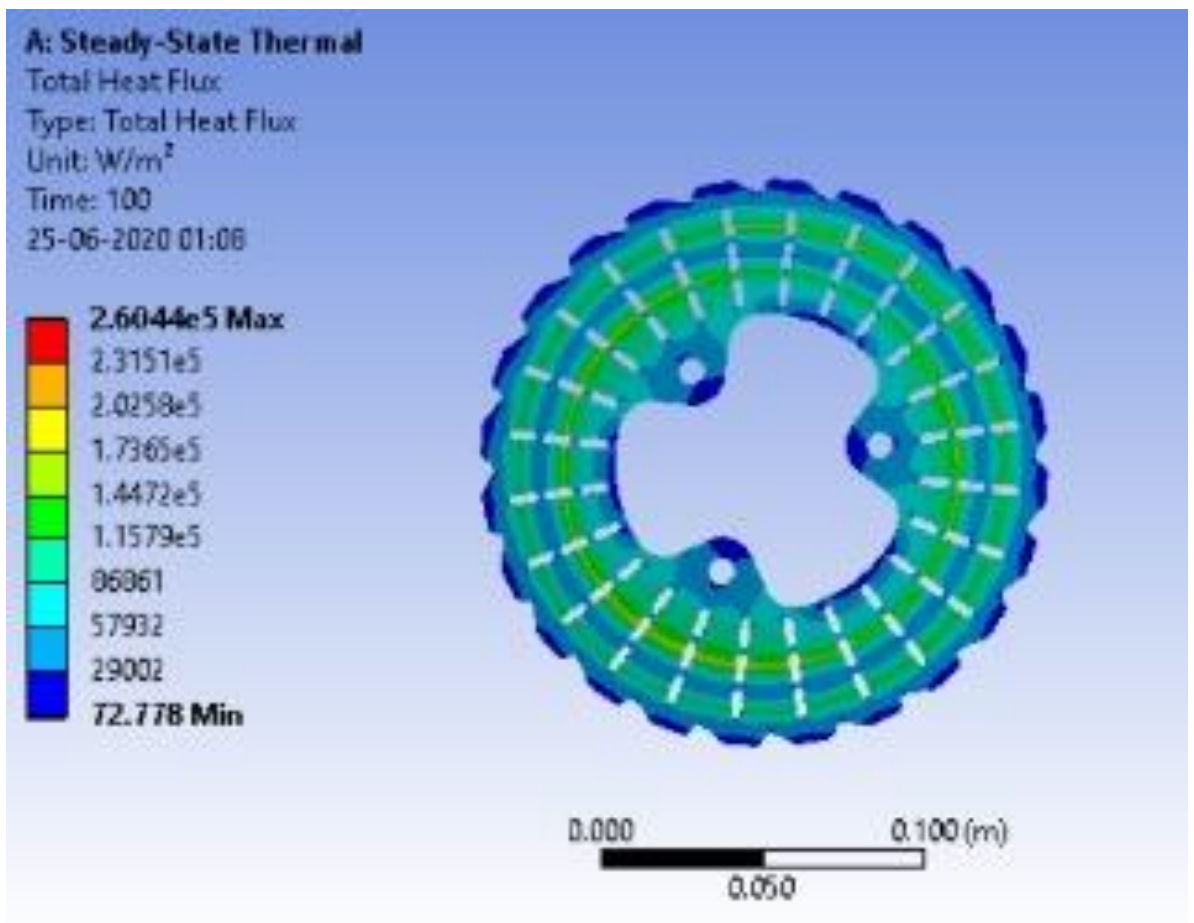

Figure 37. Total Heat Flux Contour

## 3.4 Modal Analysis

Behavior of rear disc brake has been observed in different modes.

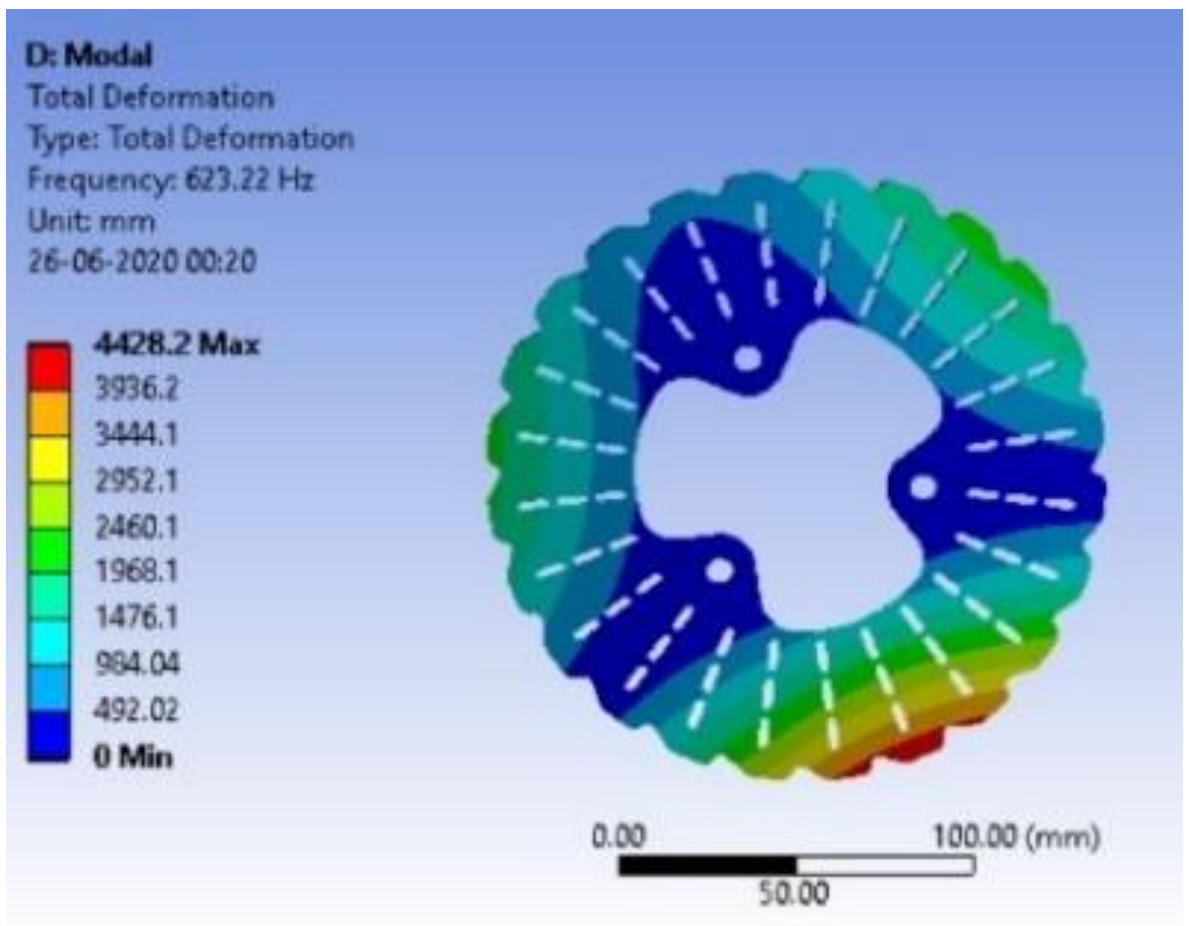

Figure 38. Total Deformation Contour for Mode 1 (Z-axis)

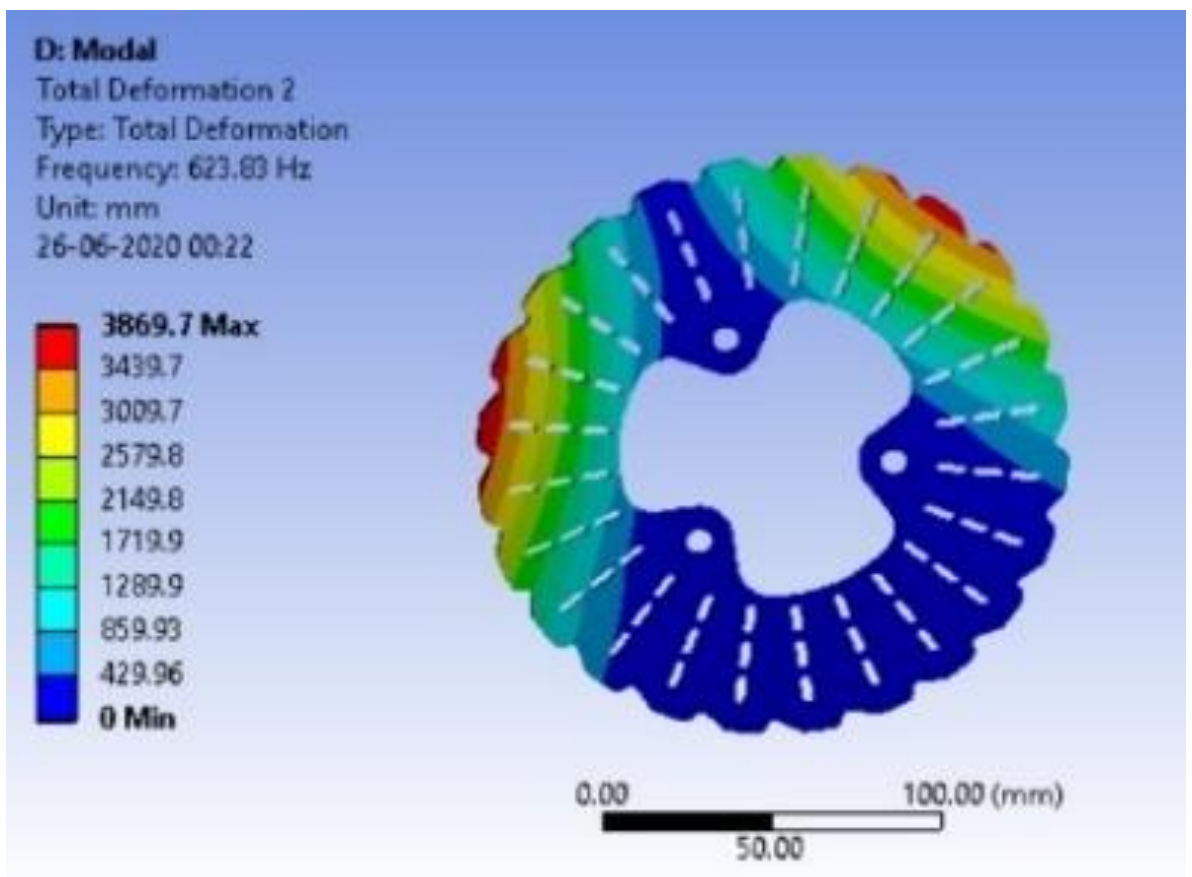

Figure 39. Total Deformation Contour for Mode 2 (Z-axis)

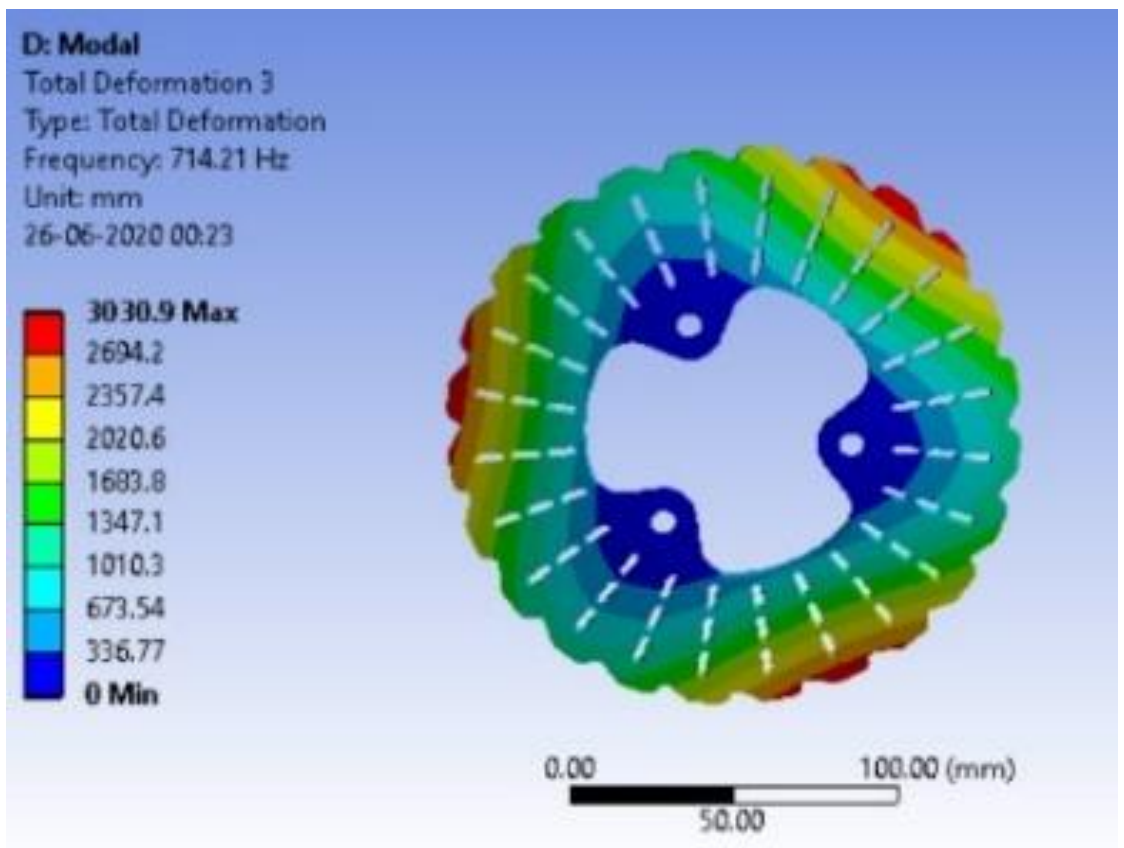

Figure 40. Total Deformation Contour for Mode 3 (Z-axis)

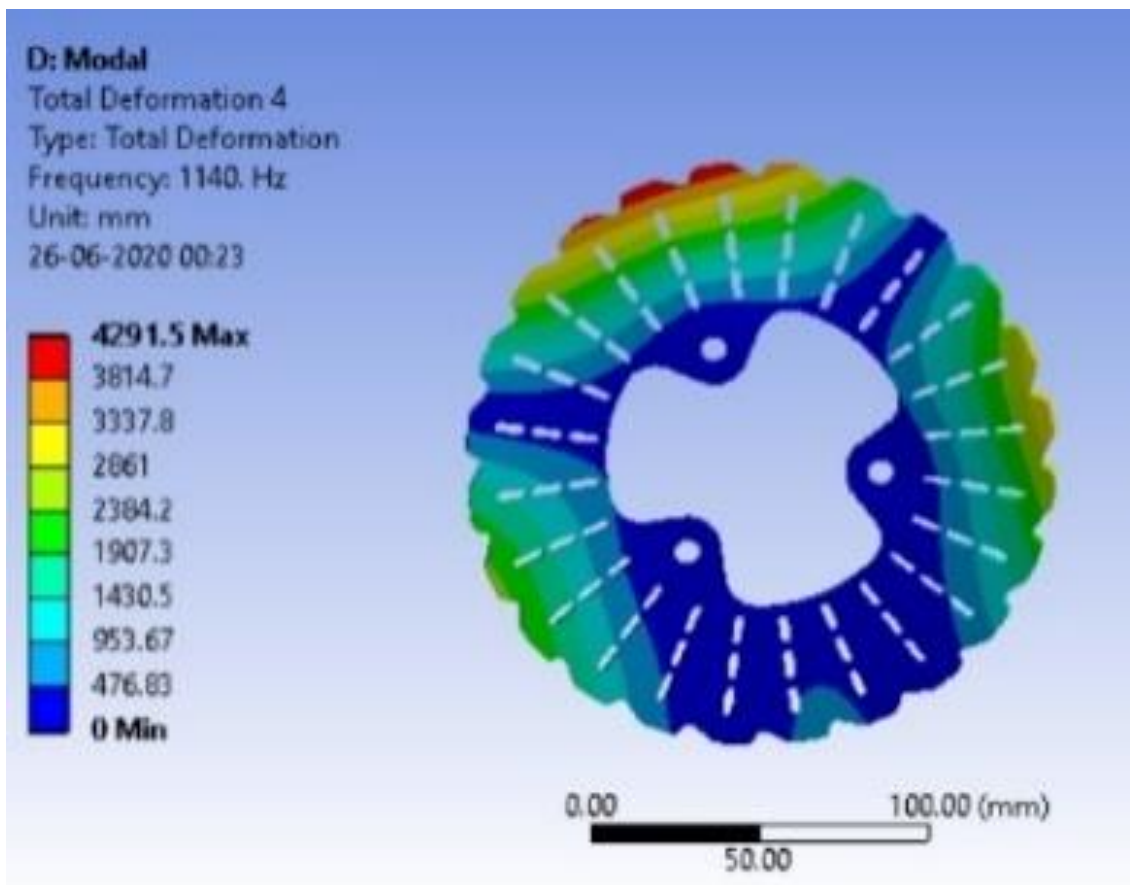

Figure 41. Total Deformation Contour for Mode 4 (Z-axis)

Table 16. Frequency with respect to different modes

| MODE | Frequency |
|---|---|
| 1 | 623.22 Hz |
| 2 | 623.83 Hz |
| 3 | 714.21 Hz |
| 4 | 1140 Hz |
| 5 | 1140.2 Hz |
| 6 | 1470.6 Hz |

## 3.5 CFD Air Simulation

Inlet Velocity, 16m/sec and Viscous laminar flow has been taken into consideration for air simulation.

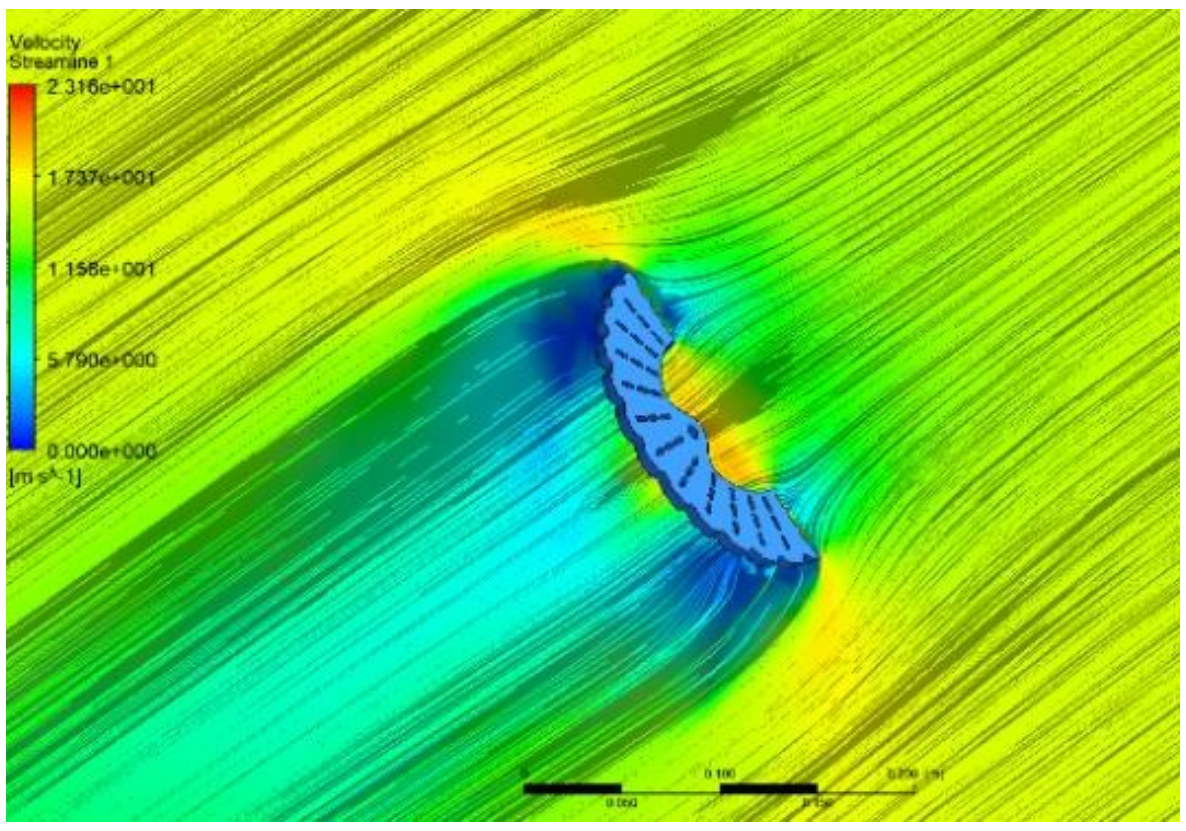

Figure 42. Velocity flow streamline

## 3.6 CFD Heat Transfer

Table 17. Meshing Results

| S.No | Parameter | Details |
|---|---|---|
| 1 | Nodes | 104657 |
| 2 | Elements | 484125 |
| 3 | Method | Fine Meshing |

Inlet Velocity, 16m/sec and heat transfer coefficient for disc brake having laminar type flow has been taken into the consideration.

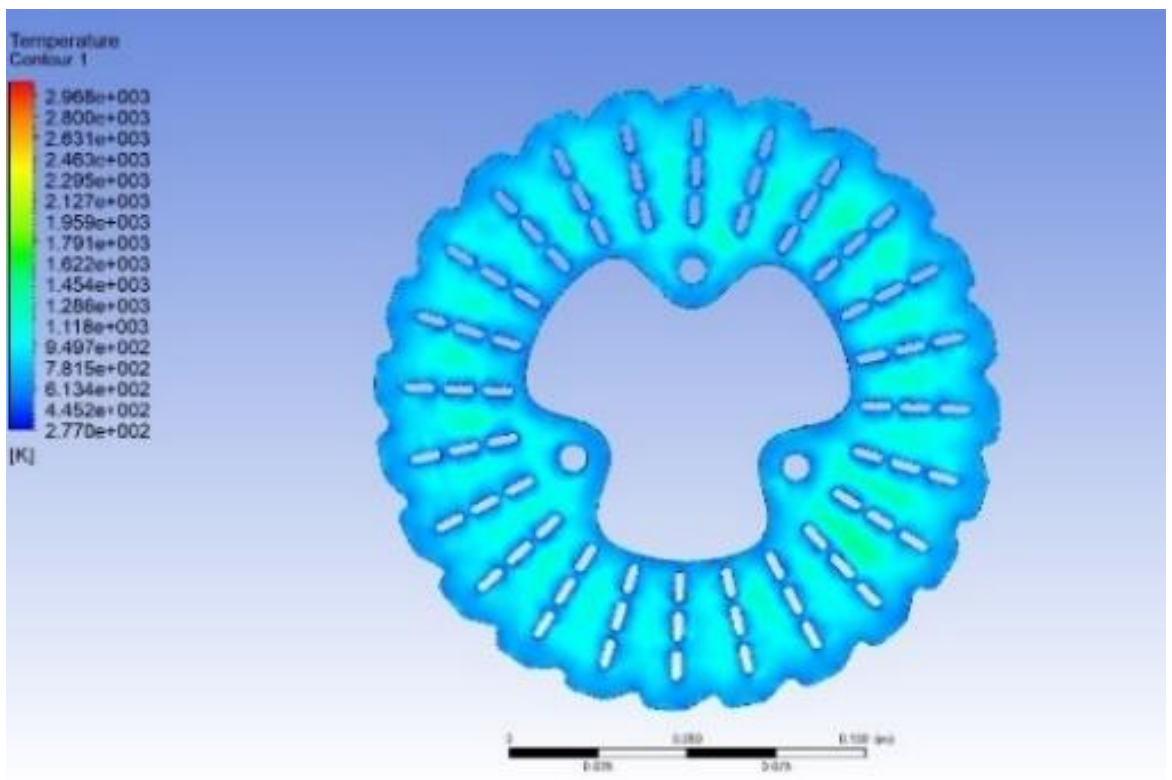

Figure 43. Face of disc brake facing velocity inlet

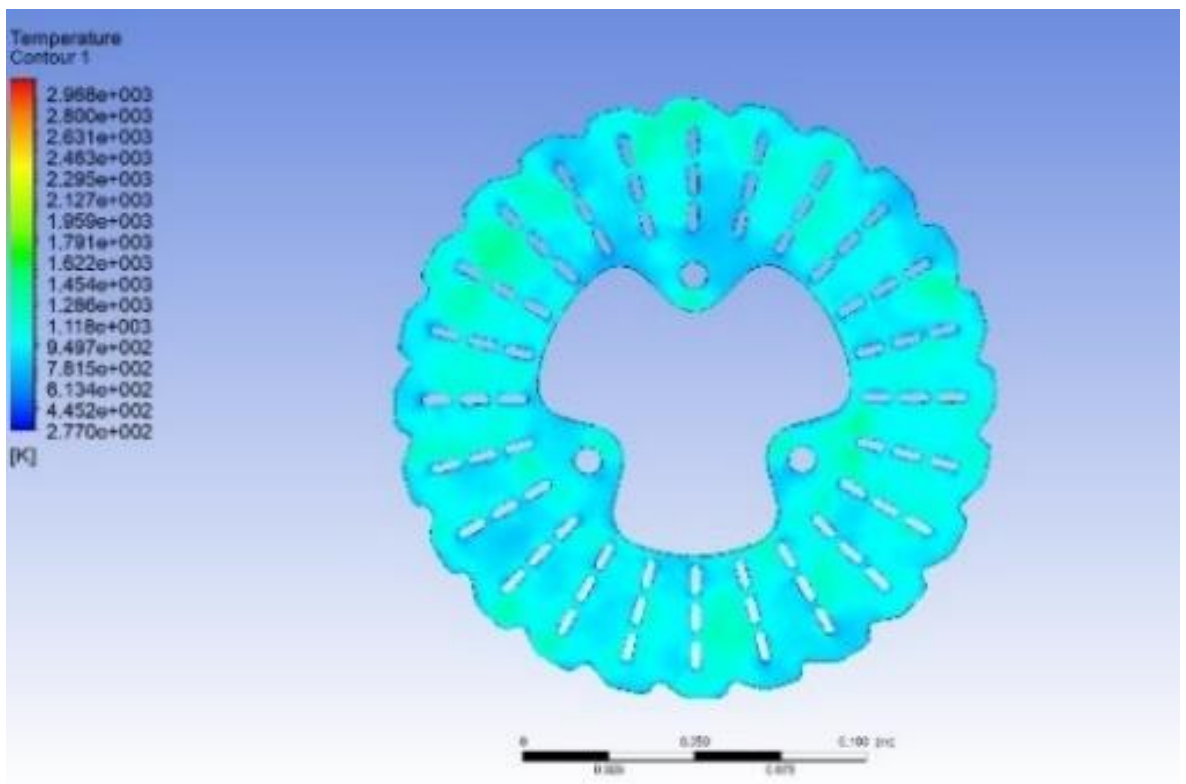

Figure 44. Face of disc brake facing velocity outlet

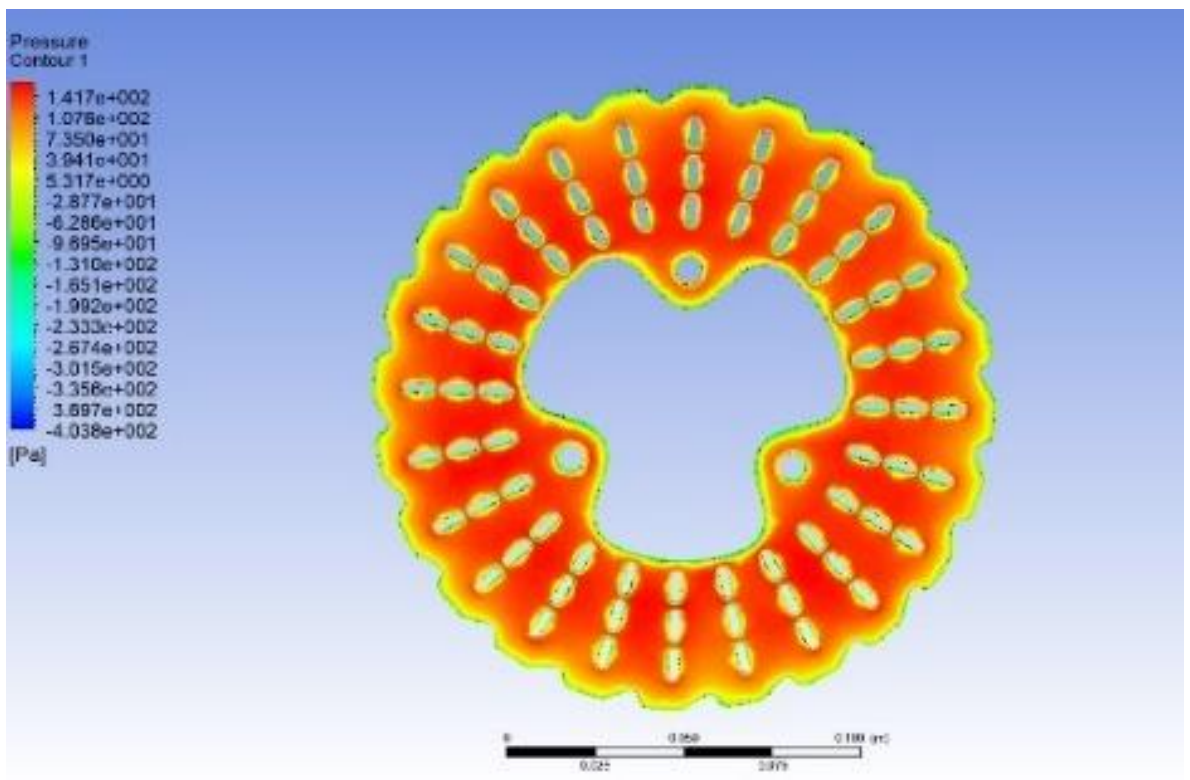

Figure 45. Face of disc brake facing pressure inlet

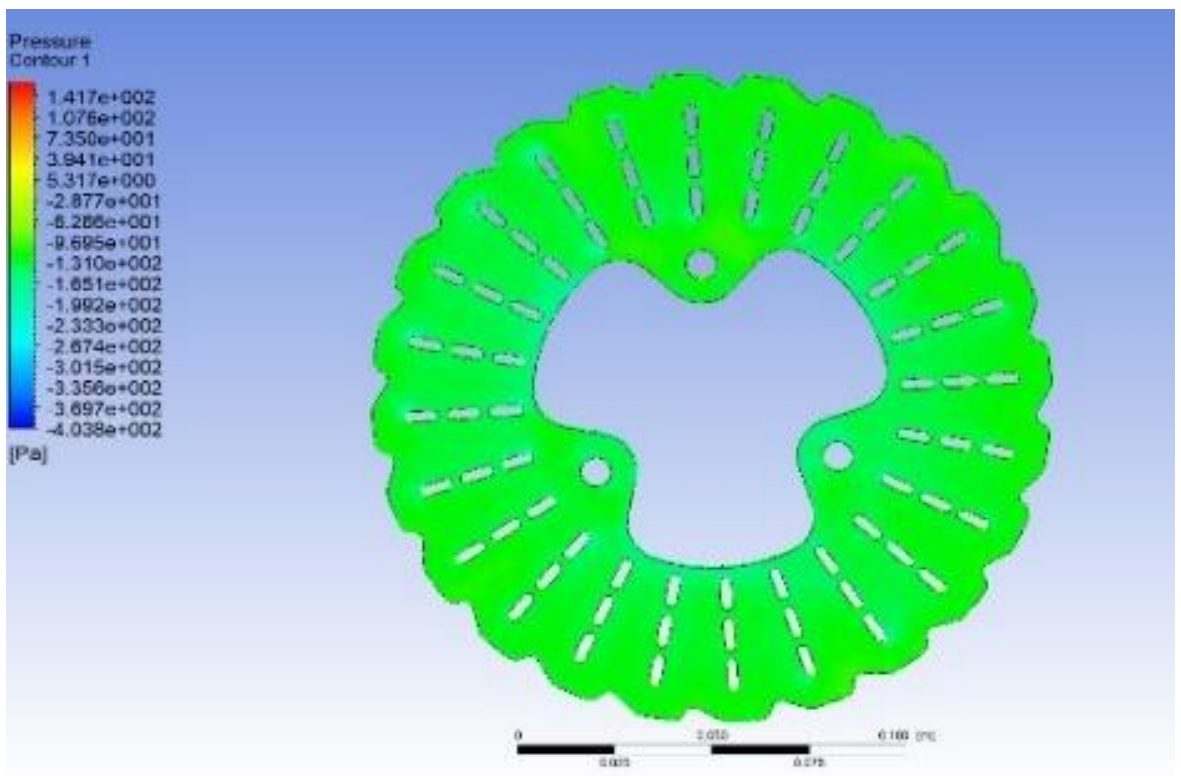

Figure 46. Face of disc brake facing pressure outlet

CFD heat transfer simulation has been carried out for to depict the behaviour of rear disc brake in terms of temperature and pressure at inlet and outlet boundaries.

## 4. Braking Efficiency

Stopping Distance = Reaction Distance + Braking Distance

Reaction Distance:

Reaction time(t) = 0.8 sec

Speed of vehicle = 16 m/s

Reaction Distance = Reaction Time * Speed of vehicle

Reaction Distance = 12.8 m

Braking Distance:

Mass of Vehicle including driver = 310 kg

Speed of vehicle = 16 m/s

Braking Force applied (F) = 13227 N

Braking Distance = {½ * m * v^ (2)} / F

Braking Distance = 3 m

Stopping Distance = Reaction Distance + Braking Distance

Stopping Distance = 15.8 m

Braking Efficiency = {v^ (2) / 2*g*Stopping distance} * 100

Braking Efficiency = 82.58 %

## 5. Summary

The mathematical model has been generated for carrying out multi-objective genetic algorithm optimization. Analysis and optimization of rear disc brake from thermal analysis point of view have been carried out for laminar heat flow convection and the behavior of the improvized and initial designed brake disc was observed in the situation of forced convection and normal convection and it has been depicted less temperature has been obtained in case of the laminar heat flow convection as compared to forced and normal convection. Computational fluid dynamics flow and heat transfer simulation has been carried out to depict the flow and temperature contours in the optimized designed disc brake respectively and simultaneously converged residual plots have been obtained using 2nd-degree order. In the computational fluid dynamics air simulation for the optimized design, we can observe the streamline flow are very smooth due to the holes and vents in the brake disc resulting in better heat dissipation as compared to the initial design. To meet the frequency of rear disc brake to firing frequency of engine, brake disc has been optimized in terms of vibration considering all the parameters. Temperature contour has helped us to depict the area for caliper mounting. Brake torque produced by the brake caliper for the rear braking system is greater than the frictional torque produced at the rear wheels and caliper force and torque are applied to the rear disc brake through the brake pad. This brake disc is going to be deployed as a common brake disc in the rear part as per the ATV architecture responsible for providing effective rear wheel locking. By optimizing the suspension & braking parameters, the proper amount of torque and force is now being acted on the brakes during bump and droop in endurance, maneuverability, and hill climb so that the vehicle brake system performs effectively in all the conditions. The below tables are demonstrating the values obtained from the computational simulation and mathematical calculations.

| S.No | Parameter | Initial Design (Value) | Final Design (Value) |
|---|---|---|---|
| 1 | Maximum Equivalent Stress | 46.523 MPa | 70.483 MPa |
| 2 | Minimum Factor of Safety | 3.8691 | 2.54 |
| 3 | 1st Order Frequency | 874.8 Hz | 623.22 Hz |
| 4 | Better Flow Simulation | No | Yes |
| 5 | Weight (in Kg) | 1.58 | 0.74 |
| 6 | (Max. Temperature) | | |
| a. | Laminar Convection | 135.84 °C | 118.85 °C |
| b. | Normal Convection | 576.91 °C | 573.45 °C |
| c. | Forced Convection | 446.02 °C | 418.91 °C |

| S.No | Parameter | Value |
|---|---|---|
| 1 | Brake Torque (Rear) | 144 N-m |
| 2 | Frictional Torque (Rear Wheels) | 92 N-m |

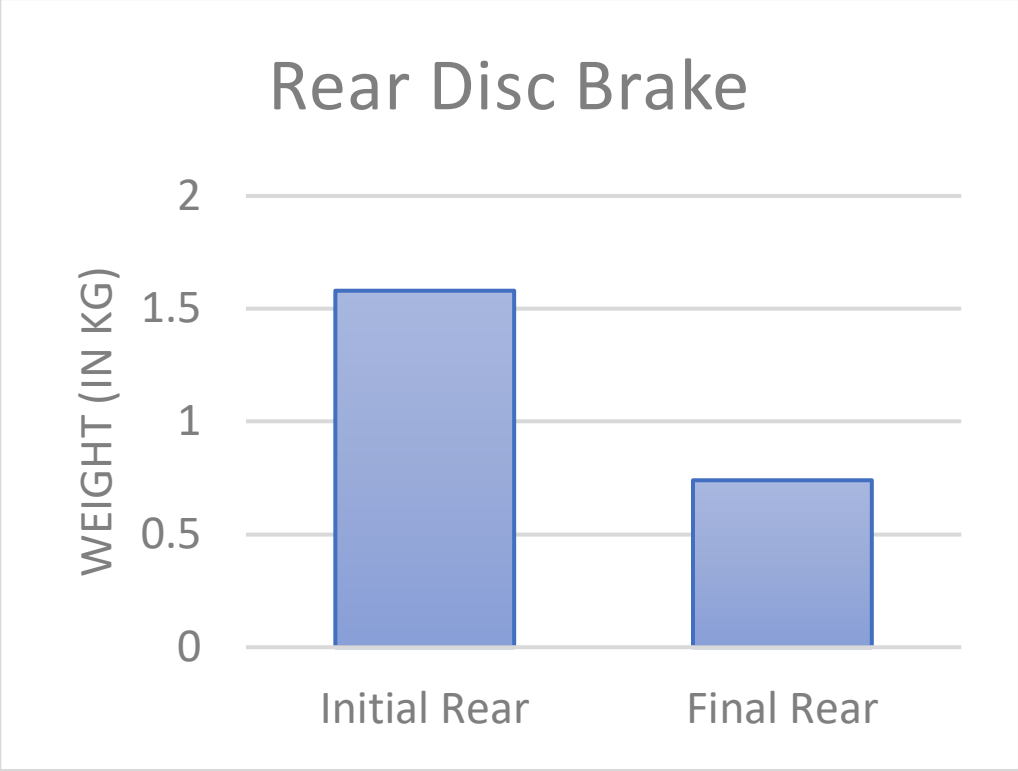

## 6. Conclusion

The newly designed rear disc brake is optimum in terms of lightweight, force, and torque bearing capacity and better thermal dissipation. The maximum temperature obtained in the computational fluid dynamics heat transfer simulation for the brake disc is 1791 K for a laminar flow and simutaneoulsy the maximum temperature obtained in the brake disc during thermal analysis is 118.85 ◦C for a laminar heat flow convection. In addition to that, the maximum temperatures obtained in the brake disc during thermal analysis for forced convection and normal convection are 418.91 ◦C and 573.45 ◦C respectively. A weight reduction of 0.84 kg has been obtained for the newly designed disc brake. The tire coefficient of friction has been considered 0.67 to incorporate all the road conditions and a braking distance of 3m has been incorporated for an off-road vehicle to obtain an effective braking efficiency of 83%. In addition to that, brake torque produced by the brake caliper for the rear braking system is greater than the frictional torque produced at the rear wheels which ensures complete locking of wheels when the brakes are applied in the vehicle.